\documentclass[prc,aps,showpacs,showkeys,groupedaddress,floatfix,superscriptaddress]{revtex4-1}
\usepackage{epsfig}
\usepackage{dcolumn}
\usepackage{bm}% bold math
\usepackage[utf8]{inputenc}
\usepackage{graphics}
\usepackage{graphicx}
\usepackage{epsfig}
\usepackage{amssymb}
\usepackage{amsmath}
\usepackage{color}
\usepackage{caption}
\usepackage{subcaption}

\begin{document}
\title{Thermal properties of a two-dimensional Duffin-Kemmer-Petiau oscillator under an external magnetic field in the presence of a minimal length}

\author{Houcine Aounallah}
\email{houcine.aounallah@univ-tebessa.dz}
\affiliation{  Laboratory of Applied and Theoretical Physics, Larbi Tebessi University, 12000 Tebessa, Algeria}

\author{Bekir Can L\"{u}tf\"{u}o\u{g}lu}
\email{Corresponding author : bclutfuoglu@akdeniz.edu.tr}

\affiliation{Department of Physics, Akdeniz University, Campus 07058 Antalya, Turkey}

\affiliation{Department of Physics,  University of Hradec Kr\'{a}lov\'{e}, Rokitansk\'{e}ho 62, 500\,03 Hradec Kr\'{a}lov\'{e}, Czechia}

\author{Jan K\v{r}\'{i}\v{z}}
\email{jan.kriz@uhk.cz}
\affiliation{Department of Physics,  University of Hradec Kr\'{a}lov\'{e}, Rokitansk\'{e}ho 62, 500\,03 Hradec Kr\'{a}lov\'{e}, Czechia}

\date{\today}
\begin{abstract}
Generalized uncertainty principle puts forward the existence of the shortest distances and/or maximum momentum at the Planck scale for consideration. In this article, we investigate the solutions of a two-dimensional Duffin-Kemmer-Petiau (DKP) oscillator within an external magnetic field in a minimal length (ML) scale. First, we obtain the eigensolutions in ordinary quantum mechanics. Then, we examine the DKP oscillator in the presence of an ML for the spin-zero and spin-one sectors. We determine an energy eigenvalue equation in both cases with the corresponding eigenfunctions in the non-relativistic limit. We show that in the ordinary quantum mechanic limit, where the ML correction vanishes, the energy eigenvalue equations become identical with the habitual quantum mechanical ones. Finally, we employ the  Euler-Mclaurin summation formula and obtain the thermodynamic functions of the DKP oscillator in the high-temperature scale.
\end{abstract}

\maketitle

\section{Introduction}

During recent years, the number of extensive research subjected to the Generalized Uncertainty Principle (GUP)  in quantum mechanics \cite{Chung_et_al_2019}, quantum electrodynamics \cite{Bosso_2018}, quantum cosmology \cite{Bosso_et_al_2019}, quantum gravity \cite{Villalpando_et_al_2019}, black-hole physics \cite{Liang_et_al_2018}, and its thermodynamics \cite{Saghafi_et_al_2017} have increased. GUP admits that absolute precision in position or momentum measurements does not exist and it predicts a lower bound fundamental scale around the Planck scale. Mathematically, the GUP concept is given by  modified commutation relations among position and momentum operators of the standard Heisenberg algebra. For the alteration, a small positive parameter, $\beta$, is employed. This minimal length (ML) parameter  modifies the algebra by
\begin{eqnarray}
\left[\mathbf{\hat{X}_{i}},\mathbf{\hat{P}_{j}}\right]&=&i\hbar\delta_{ij}\big(1+\beta p^{2}\big),\label{C1}\\
\left[\mathbf{\hat{X}_{i}},\mathbf{\hat{X}_{j}}\right]&=&-2i\hbar\beta\big(1+\beta p^{2}\big)\varepsilon_{ijk}\hat{L}_{k}, \label{C2}\\
\left[\mathbf{\hat{P}_{i}},\mathbf{\hat{P}_{j}}\right]&=&0, \label{C3}
\end{eqnarray}
where $\hat{L}_{k}$ is the usual angular momentum operator that satisfies the standard Heisenberg algebra. As a consequence of this modification, a more general expression, namely GUP,
\begin{eqnarray}
  \left(\triangle X\right)\left(\triangle P\right)\geq i\hbar\bigg(1+\beta\big(\triangle P\big)^{2}\bigg),
\end{eqnarray}
is obtained instead of the Heisenberg uncertainty relation  \cite {1, 2, 3, 4, Merad_et_al_2012}.  Note that when the ML parameter goes to zero, the GUP expression turns to the standard uncertainty relationship in the ordinary quantum mechanics (OQM).

In the past hundred years, exact or approximate solutions to the Schr\"odinger equation have been investigated to describe the physical processes in the OQM \cite{Badalov_et_al_2019, Li_et_al_2019, Lutfuoglu_et_al_2016}. This non-relativistic equation has also been examined within the GUP perspective \cite{Hassanabadi_et_al_2013, Hassanabadi_et_al_2014, Haouat_2014, Bhat_et_al_2017, Khorram_et_al_2019}. In particular, Bhat \emph{et al.} calculated the correction term to the energy eigenvalue function of a gravitational quantum well, by employing infinite extra dimensions \cite{Bhat_et_al_2017}.

The Klein-Gordon(KG) and the Dirac equations were frequently used in studies examining relativistic physical processes \cite{Ikot_et_al_2016, Lutfuoglu_2019, Hosseini_et_al_2019, Arda_et_al_2019, Oliveira_et_al_2019, Yesiltas_2019}. These well-known differential equations were also explored within the GUP.  For example, in the presence of the ML, Jana \emph{et al.} obtained an exact solution to the KG equation by employing a linear and vector potential energy \cite{Jana_et_al_2009}. In 2018, Boumali \emph{et al.} investigated a two dimensional KG oscillator under the GUP \cite{Boumali_et_al_2018}.  In another study, Elviyanti \emph{et al.} employed Asymptotic Iteration Method to examine the Hult\'en potential energy in the KG equation in the ML \cite{Elviyanti_et_al_2018}. Additionally, in 2013, Menculini \emph{et al.} studied the relativistic Landau problem in Dirac equation within the presence of the ML and reported an exact solution of the wave function in the momentum space \cite{Menculini_et_al_2013}. In 2015, Pedram \emph{et al.} studied the two dimensional Dirac equation by employing a non-varying magnetic field in an ML. They concluded that the solution that was found by  Menculini \emph{et al.} was a subset of the general solution which was correlated with the even quantum numbers \cite{Pedram_et_al_2015}.  Ikot \emph{et al.} examined the two-dimensional harmonic oscillator problem within the framework of ML on the Dirac equation in the commutative and noncommutative space, respectively \cite{Ikot_et_al_2015}. In a very recent paper, Hamil \emph{et al.} investigated the Dirac oscillator in two and three dimensions. They announced that their results could predict the upper bound of the GUP parameter of the relativistic Landau levels in graphene \cite{Hamil_et_al_2019}.

Duffin-Kemmer-Petiau (DKP) equation is another relativistic equation that appeared a century ago \cite {23,24,25,26}. It is a Dirac-like first-order equation that describes the dynamics of a spin-zero and spin-one boson together. At the same time, it possesses a very complex algebraic structure.  Moreover, before the 1970's its solutions were considered to be equivalent to the KG and Proca equations in on-shell cases.   However, after then, it was comprehended that in special cases that involve a symmetry breaking and hadronic processes,  the solution of the DKP equation differs from those of other equations \cite{31}. Recently, in the literature, we have seen comprehensive studies that focused on the solution of the DKP equation. For example, Boumali \emph{et al.} investigated a two dimensional DKP oscillator under the effect of a magnetic field \cite{31}. Falek \emph{et al.} investigated a three dimensional DKP oscillator  \cite{Falek_et_al_2010}. Moreover, they considered  a deformed DKP oscillator with Snyder-de Sitter algebra in a momentum space with and without a magnetic field \cite{Falek_et_al_2019, Falek_et_al_2017}. One of the authors of this article, Aounallah with his collaborator, obtained exact solutions of the DKP equation with Aharonov-Bohm and Coulomb potential energies in the commutative \cite{Houcini_et_al_2018} and non commutative \cite{Houcini_et_al_2019} space-times which are assumed to be produced by a cosmic string. Hosseinpour  \emph{et al.} discussed the dynamic of the DKP particles in the space-time that are initiated with a spinning cosmic string \cite{Hosseinpour_et_al_2019}. Wang \emph{et al.} studied the spin-one sector of the DKP oscillator in non commutative space in two dimension \cite{Wang_ek_2017, Wang_et_al_2018}. Hun \emph{et al.} considered the DKP oscillator in a curved space-time \cite{Hun_et_al_2019}. In addition to these studies, Lunardi and Chargui contributed the field with remarkable articles. In particular, Lunardi, he discussed the equivalency of the spin-one and spin-zero representation of the DKP equation in one dimension \cite{Lunardi_2017}. Chargui revisited the DKP equation with linear potential energies in one dimension and reported some inaccurate discussions in the literature \cite{Chargui_2018}.

Comprehending the thermal properties of physical systems is important for their applications in the real world. We observe that scientists have begun to discuss the thermal properties of the models they are handling \cite{{01, 02, 03}}. One of the pioneering work in this area was given by Pacheco \emph{et al.}  in 2003 \cite{Pacheco_et_al_2003}. There, they investigated the thermodynamic properties of a one-dimensional Dirac oscillator. Later, they extended their work to three dimensions \cite{Pacheco_et_al_2014}. Nouicer analyzed a one-dimensional Dirac oscillator in the presence of the ML and examined its statistical properties \cite{Nouicer_2006}. Hamil \emph{et al.} studied the Dirac and KG oscillators and reported their thermal properties on anti-de Sitter space \cite{Hamil_et_al_2018}. Wu \emph{et al.} investigated thermodynamic functions of a two-dimensional DKP oscillator in the GUP formalism \cite{Wu_et_al_2017}.

The presence of an external magnetic field plays a crucial role in physical phenomena. For example,  the energy levels of hydrogen-like atoms are splitting with the external magnetic field and this fact is known as the Zeeman effect. On the other hand, two dimensional systems such as graphene,  gather interest, especially in the nano-physics sector.
The effects of the external magnetic field and the presence of the ML on the thermal quantities in a two-dimensional Dirac oscillator are explored \cite{04,05}. To our knowledge,  the solutions of a two-dimensional DKP oscillator under an external magnetic field in the presence of the ML has not been discussed with the statistical point of view.  Our main motivation is to obtain the modification of the thermodynamic functions. We expect that the introduction of the external magnetic field with the existing of the ML will present important consequences on the thermal properties of the DKP oscillator. We believe this investigation is going to fill the missing gap in the literature.

We organize the article as follows. In sect. \ref{FDKP}, we review the irreducible representations of the DKP equation for both the spin-zero and spin-one particles. In sect. \ref{oQMDKP}, we derive an exact  solution to the two-dimensional DKP oscillator under the effect of an homogeneous magnetic field by employing the polar coordinates in  the  momentum  space within  the rules of ordinary quantum  mechanics (OQM). In sect. \ref{MLQMDKP}, we investigate an exact solution of the problem in the GUP in the non-relativistic limit. In sect. \ref{TP} we obtain the thermodynamic functions by employing Euler-Mclaurin summation formula in the high temperature approach for the spin-zero DKP oscillator in the canonical ensemble formalism. Moreover, we demonstrate the thermal properties in the figures.  We conclude the article in sect. \ref{Concl}.

\section{Formalism of the DKP equation} \label{FDKP}

In flat space-time, the DKP equation is given by \cite{23,24,25,26}
\begin{eqnarray}
\big(i \hbar \beta^{\mu}\partial_{\mu}-Mc\big)\Psi(\vec{r},t)=0, \hspace{1cm} \mu=0,1,2,3, \label{FreeDKP}
\end{eqnarray}
where $M$, $\hbar$ and $c$ denote the mass,  reduced Planck constant, and speed of light,  respectively.  $\beta^{\mu}$ matrices satisfy the DKP algebra \cite{26}
\begin{eqnarray}
\beta^{\kappa}\beta^{\nu}\beta^{\lambda}+\beta^{\lambda}\beta^{\nu}\beta^{\kappa}=g^{\kappa\nu}\beta^{\lambda}+g^{\nu\lambda}\beta^{\kappa}.  \label{DKPalgebra}
\end{eqnarray}%
Here, $\kappa, \nu, \lambda=0,1,2,3$. We examine the problem in the Minkowski space-time and we use the metric tensor, $g^{\mu\nu}$, with the  $\text{diag}\left(1,-1,-1,-1\right)$ signature. Note that there are $126$ independent elements in the algebra. These elements can be reduced into three irreducible representations of dimensions one, five, and ten. Among them, one-dimensional representation is the trivial one. The others, namely the five and ten-dimensional representations, describe spin-zero and spin-one particle dynamics, respectively.

For a scalar particle, we choose the representation that is given with  $ 5\times5 $ matrices \cite{31}
\begin{eqnarray}
\beta^{0}&=&\left(\begin{array}{cc}
\mathbf{\rho}_{2\times 2} & \mathbf{\tilde{0}}_{2\times3}\\
\mathbf{\tilde{0}}^{T}_{3\times2} & \mathbf{\overline{0}}_{3\times3}
\end{array}\right),\\
\beta^{i}&=&\left(\begin{array}{cc}
\mathbf{\hat{0}}_{2\times 2} & \mathbf{\rho}^{i}_{2\times 3}\\
-\mathbf{\rho}^{i^{T}}_{3\times2} & \mathbf{\overline{0}}_{3\times3}
\end{array}\right),
\end{eqnarray}%
where $i=1,2,3$. Note that the subscripts indicate the row and column numbers. $\mathbf{\hat{0}}_{2\times 2},\mathbf{\tilde{0}}_{2\times3},\mathbf{\overline{0}}_{3\times3}$ are zero matrices, while the non-zero matrices are \cite{31}
\begin{eqnarray}
 \rho_{2\times 2}&=&\left(\begin{array}{cc}
0 & 1\\
1 & 0
\end{array}\right),\\
\rho^{1}_{2\times 3}=\left(\begin{array}{ccc}
-1 & 0 & 0\\
0 & 0 & 0
\end{array}\right), \,\,\,\,\,\,
\rho^{2}_{2\times 3}&=&\left(\begin{array}{ccc}
0 & -1 & 0\\
0 & 0 & 0
\end{array}\right),\,\,\,\,\,\,
\rho^{3}_{2\times 3}=\left(\begin{array}{ccc}
0 & 0 & -1\\
0 & 0 & 0
\end{array}\right).
\end{eqnarray}
For a vector particle, we take the spin-one representation that is given with  $ 10\times10 $ matrices \cite{31}
\begin{eqnarray}
\beta^{0}&=&\left(\begin{array}{cccc}
\mathbf{\overline{0}}_{3\times3} & \mathbf{\overline{0}}_{3\times3} & -\mathbf{I}_{3\times3} & \mathbf{\check{0}}^{T}_{3\times1}\\
\mathbf{\overline{0}}_{3\times3} & \mathbf{\overline{0}}_{3\times3} & \mathbf{\overline{0}}_{3\times3} & \mathbf{\check{0}}^{T}_{3\times1}\\
-\mathbf{I}_{3\times3} & \mathbf{\overline{0}}_{3\times3} & \mathbf{\overline{0}}_{3\times3} & \mathbf{\check{0}}_{3\times1}^{T}\\
\mathbf{\check{0}}_{1\times3} & \mathbf{\check{0}}_{1\times3} & \mathbf{\check{0}}_{1\times3} & 0
\end{array}\right), \\
\beta^{k}&=&\left(\begin{array}{cccc}
\mathbf{\overline{0}}_{3\times3} & \mathbf{\overline{0}}_{3\times3} & \mathbf{\overline{0}}_{3\times3} & i\mathbf{K}^{j^{T}}\\
\mathbf{\overline{0}}_{3\times3} & \mathbf{\overline{0}}_{3\times3} & \mathbf{s}^{j}_{3\times3} & \mathbf{\check{0}}_{3\times1}^{T}\\
\mathbf{\overline{0}}_{3\times3} & -\mathbf{s}_{3\times3}^{j} & \mathbf{\overline{0}}_{3\times3} & \mathbf{\check{0}}_{3\times1}^{T}\\
i\mathbf{K}^{j} & \mathbf{\check{0}}_{1\times3} & \mathbf{\check{0}}_{1\times3} & 0
\end{array}\right),
\end{eqnarray}%
where $ j=1,2,3 $, and
\begin{equation}
\begin{array}{cc}
\mathbf{\check{0}}_{3\times1}^T=\left(\begin{array}{c}
0 \\
0 \\
0
\end{array}\right), & \mathbf{I}=\end{array}\left(\begin{array}{ccc}
1 & 0 & 0\\
0 & 1 & 0\\
0 & 0 & 1
\end{array}\right),
\end{equation}%
\begin{equation}
\begin{array}{ccc}
\mathbf{s}^{1}=i\left(\begin{array}{ccc}
0 & 0 & 0\\
0 & 0 & -1\\
0 & 1 & 0
\end{array}\right), & \mathbf{s}^{2}=i\left(\begin{array}{ccc}
0 & 0 & 1\\
0 & 0 & 0\\
-1 & 0 & 0
\end{array}\right), & \mathbf{s}^{3}=i\left(\begin{array}{ccc}
0 & -1 & 0\\
1 & 0 & 0\\
0 & 0 & 0
\end{array}\right),\end{array}
\end{equation}%
\begin{equation}
\begin{array}{ccc}
\mathbf{K}^{1}=\left(\begin{array}{ccc}
1 & 0 & 0\end{array}\right), & \mathbf{K}^{2}=\left(\begin{array}{ccc}
0 & 1 & 0\end{array}\right), & \mathbf{K}^{3}=\left(\begin{array}{ccc}
0 & 0 & 1\end{array}\right).\end{array}
\end{equation}%

\section{DKP oscillator within OQM} \label{oQMDKP}
In this section, we formulate the DKP oscillator and investigate its solution for spin-zero and spin-one particles in OQM limit. We assume that the potential energy and external magnetic field that are time-independent.

\subsection{Case of the spin-zero particle} \label{DKP-0}

In $(2+1)$ dimension, we express a DKP oscillator for a spin-zero particle with a non-minimal coupling to an external magnetic field as \cite{31}
\begin{eqnarray}
\bigg[\beta^{0}E-c \beta^{1} \Big(\mathbf{p_{x}}-\frac{eA_{x}}{c}-i M \omega \eta^0 \mathbf{x}\Big)-c \beta^{2}\Big(\mathbf{p_{y}}-\frac{eA_{y}}{c}-iM\omega \eta^0 \mathbf{y} \Big)-Mc^{2}\mathbb{1}\bigg]\varPsi=0. \label{DKPspin0}
\end{eqnarray}%
Here, $ E $ is the energy eigenvalues and $\omega$ is the frequency of the oscillator. We denote the momentum and position operators with bold letters. The matrix $\eta^0$ is defined by $\eta^0\equiv 2 (\beta^0)^2-\mathbb{1}$ and its square is equal to the identity matrix. We employ a vector potential with two components: $ A_{x}=-\frac{B}{2}y $ and $ A_{y}=\frac{B}{2}x$. Here, we symbolize the strength of the magnetic field via $B$.  We express the transposed form of the spatial wave function with
\begin{eqnarray}
\varPsi^{T} \equiv \left(\begin{array}{ccccc}
\varPsi_{1} & \varPsi_{2} & \varPsi_{3} & \varPsi_{4} & \varPsi_{5}\end{array}\right). \label{PsiTspin0}
\end{eqnarray}%
We substitute Eq.  (\ref{PsiTspin0}) into Eq. (\ref{DKPspin0}) and obtain the following coupled equations.
\begin{eqnarray}
-Mc^{2}\varPsi_{1}+E\varPsi_{2}+c\big(\mathbf{p_{x}}-M\widetilde{\omega}\mathbf{y}+iM\omega\mathbf{x}\big)\varPsi_{3}+c\big(\mathbf{p_{y}}+M\widetilde{\omega}\mathbf{x}+iM\omega \mathbf{y}\big) \varPsi_{4} &=& 0, \label{spin-0_eq_1} \\
E\varPsi_{1}-Mc^{2}\varPsi_{2}&=&0, \label{spin-0_eq_2}\\
c\big(\mathbf{p_{x}}-M\widetilde{\omega}\mathbf{y}-iM\omega \mathbf{x}\big)\varPsi_{1}+Mc^{2}\varPsi_{3}&=&0, \label{spin-0_eq_3}\\
c\big(\mathbf{p_{y}}+M\widetilde{\omega}\mathbf{x}-im_{0}\omega \mathbf{y}\big) \varPsi_{1} +Mc^{2}\varPsi_{4}&=&0, \label{spin-0_eq_4}\\
\varPsi_{5}&=&0,  \label{spin-0_eq_5}
\end{eqnarray}%
where  $ \widetilde{\omega}$ is the cyclotron frequency and defined by $\frac{\left|e\right|B}{2Mc}$. Note that the fifth component of the wave equation is equal to zero. By writing the second, third and fourth components in terms of the first component, we obtain
\begin{eqnarray}
&&\bigg[c^2\Big(\mathbf{p_{x}}-M\widetilde{\omega}\mathbf{y}+iM\omega \mathbf{x}\Big)\Big(\mathbf{p_{x}}-M\widetilde{\omega}\mathbf{y}-iM\omega \mathbf{x}\Big)\nonumber \\&+&
c^2\Big(\mathbf{p_{y}}+M\widetilde{\omega}\mathbf{x}+iM\omega \mathbf{y}\Big)\Big(\mathbf{p_{y}}+M\widetilde{\omega}\mathbf{x}-iM\omega \mathbf{y}\Big) +\Big(M^2c^{4}-E^{2}\Big)\bigg]\varPsi_{1}=0. \label{S0main}
\end{eqnarray}%
Then, as done in \cite{Wu_et_al_2017}, we employ the operators defined in polar coordinates in the momentum space in Eq. (\ref{S0main})
\begin{eqnarray}
\mathbf{x}     &\equiv& i\hbar\bigg(\cos{\theta}\frac{\partial}{\partial p}-\frac{\sin{\theta}}{p}\frac{\partial}{\partial \theta}\bigg), \label{xop}\\
\mathbf{y}     &\equiv& i\hbar\bigg(\sin{\theta}\frac{\partial}{\partial p}+\frac{\cos{\theta}}{p}\frac{\partial}{\partial \theta}\bigg), \label{yop}\\
\mathbf{p_{x}} &\equiv& p \cos{\theta},\label{pox}\\
\mathbf{p_{y}} &\equiv& p \sin{\theta}.\label{poy}
\end{eqnarray}
We obtain
\begin{eqnarray}
\bigg[p^{2}-2M \hbar \bigg( \omega + i \widetilde{\omega} \frac{\partial}{\partial\theta}  \bigg)- M^2 \hbar^2 \Omega^2  \bigg(\frac{\partial^{2}}{\partial p^{2}}+\frac{1}{p}\frac{\partial}{\partial p}+ \frac{1}{p^{2}} \frac{\partial^{2}}{\partial\theta^{2}}\bigg) -\varsigma \bigg]\varPsi_{1}=0. \label{DKPspin0-1}
\end{eqnarray}%
where
\begin{eqnarray}
\Omega^{2}&\equiv& \widetilde{\omega}^{2}+\omega^{2}, \\
\varsigma&\equiv& \frac{E^{2}-M^{2}c^{4}}{c^{2}}. \end{eqnarray}%
Then, we separate the wave function into the spatial and angular parts
\begin{eqnarray}
\varPsi_{1}(p,\theta)&\equiv& f(p)e^{im\theta}, \label{Spat_angular}
\end{eqnarray}%
where $ m=0,\pm1,\pm2,\pm3,\cdots $. After we substitute the decomposed wave function into Eq. (\ref{DKPspin0-1}), we find
\begin{eqnarray}
\bigg[\frac{d^{2}}{dp^{2}}+\frac{1}{p}\frac{d}{dp}-\frac{m^{2}}{p^{2}}+\Big(\kappa_m^{2}-k^{2}p^{2}\Big)\bigg]f(p)&=&0. \label{DKPspin0-2}
\end{eqnarray}%
Here,
\begin{eqnarray}
\kappa_m^{2}&\equiv&\frac{2 M \hbar (\omega-m\widetilde{\omega})+\varsigma}{M^2\hbar^2 \Omega^2 },\\
k^{2}&\equiv&\frac{1}{M^2\hbar^2 \Omega^2 }.
\end{eqnarray}%
We consider the following Ansatz
\begin{eqnarray}
f(p)&\equiv& p^{|m|} e^{-\frac{k}{2}p^{2}} F(p), \label{ansatz1}
\end{eqnarray}%
and then we employ it in Eq. (\ref{DKPspin0-2}). We get
\begin{eqnarray}
F^{''}(p)+\bigg(\frac{2{|m|}+1}{p}-2kp\bigg)F^{'}(p)-\bigg(2k\Big({|m|}+1\Big)-\kappa_m^{2}\bigg)F(p)=0.  \label{DKPspin0-3}
\end{eqnarray}%
We introduce a new transformation,  namely $ t=kp^{2} $, and use in Eq.  (\ref{DKPspin0-3}). We obtain
\begin{eqnarray}
t\frac{d^{2}F(t)}{dt^{2}}+\bigg( {|m|}+1-t\bigg) \frac{dF(t)}{dt}-\bigg( \frac{1}{2}\Big({|m|}+1\Big)-\frac{\kappa_m^{2}}{4k}\bigg) F(t)=0,
\end{eqnarray}%
which is known as the Kummer's differential equation  and its solutions are expressed in terms of confluent hypergeometric functions \cite{Abramowitz_et_al_Book}.
\begin{eqnarray}
F(t)&=&N_1 \,_1F_1\bigg(\frac{1}{2}\Big({|m|}+1\Big)-\frac{\kappa_m^{2}}{4k},{|m|}+1,t\bigg)+N_2 \,_1U_1\bigg(\frac{1}{2}\Big({|m|}+1\Big)-\frac{\kappa_m^{2}}{4k},{|m|}+1,t\bigg).\,\,\,\,\,\,\,\,\,
\end{eqnarray}
Here,  $\,_1F_1\bigg(\frac{1}{2}\Big({|m|}+1\Big)-\frac{\kappa_m^{2}}{4k},{|m|}+1,t\bigg)$ and $\,_1U_1\bigg(\frac{1}{2}\Big({|m|}+1\Big)-\frac{\kappa_m^{2}}{4k},{|m|}+1,t\bigg)$  are the first and second kind confluent hypergeometric function. Note that we denote the normalization constants with $N_1$ and $N_2$. To avoid a singularity at $p=0$, thus at $t=0$, we pick $N_2=0$. Then, we adopt the condition for obtaining a polynomial type solution from confluent hypergeometric function
\begin{eqnarray}
\frac{1}{2}\Big(|m|+1\Big)-\frac{\kappa_m^{2}}{4k}&=&-n,
\end{eqnarray}
where $n=0,1,2,3,\cdots$. We derive the energy spectrum function as
\begin{eqnarray}
E_{n, m}=\pm Mc^{2}\bigg[1+\frac{2\hbar\big(m\widetilde{\omega}-\omega\big)}{Mc^{2}}+\frac{\hbar\Omega\big(4n+2({|m|}+1)\big)}{Mc^{2}}\bigg]^{1/2}. \label{Energy_spin0}
\end{eqnarray}%
This result is in agreement with \cite{31}. Finally, we express the first component of the wave function as follows
\begin{eqnarray}
{\varPsi_{1}}_{n, m} \left(p,\theta\right)=C_{n,m}p^{{|m|}}e^{-\frac{k}{2}p^{2}}\,_{1}F_{1}\left(-n;{|m|}+1;kp^{2}\right)e^{im\theta}.
\end{eqnarray}%
where $C_{n,m}$ is the normalization constant. The general solution of the first component of the wave function is the linear combination of separable solutions.
\begin{eqnarray}
\varPsi_{1}(p,\theta,t)&=& \sum_{n=0}^{\infty} \sum_{m=-n}^{n}  {\varPsi_{1}}_{n, m}(p,\theta) \exp{\bigg(-\frac{i E_{n,m}t}{\hbar}\bigg)}.
\end{eqnarray}%
\subsubsection{The non-relativistic limit}
In the non-relativistic limit we assume that the rest mass energy is much greater then the non-relativistic energy, $E^{NRL}$. We use $ E \cong Mc^2+E^{NRL}$  to derive the non-relativistic energy expression \cite{Falek_et_al_2019}. We get
\begin{eqnarray}
 E^{NRL}\cong \frac{E^2-M^2c^4}{2 Mc^2}.
\end{eqnarray}
Following a simple algebra, we obtain the non-relativistic energy as follows:
\begin{eqnarray}
 E^{NRL}_{n, m}&=& \hbar\Big[\sqrt{{\widetilde{\omega}}^2+\omega^2}\big(2n+{|m|}+1\big)+\big(m\widetilde{\omega}-\omega\big)\Big].
\end{eqnarray}
This result corrects the typo of Eq. (48) in \cite{31}.

\subsection{Case of the spin-one particle} \label{DKP-1}
In the spin-one case, unlike the spin-zero case, we use another non-trivial irreducible representation of the DKP algebra where $\beta$ matrices are defined with $ 10 \times 10 $ matrices. Therefore, the total wave function has ten components. We define the transposed form of the stationary wave function with
\begin{eqnarray}
\varPsi^{T}\equiv\left(\begin{array}{cccccccccc}
\varPsi_{1} & \varPsi_{2} & \varPsi_{3} & \varPsi_{4} & \varPsi_{5} & \varPsi_{6} & \varPsi_{7} & \varPsi_{8} & \varPsi_{9} & \varPsi_{10}\end{array}\right).
\end{eqnarray}%
We substitute the wave function in Eq. (\ref{DKPspin0}) and obtain the following coupled equations.
\begin{eqnarray}
Mc^{2}\varPsi_{1}+E\varPsi_{7}+ic\big(\mathbf{p_{x}}-M\widetilde{\omega}\mathbf{y}+iM\omega \mathbf{x}\big)\varPsi_{10}&=&0, \label{spin1-1} \\
Mc^{2}\varPsi_{2}+E\varPsi_{8}+ic\big(\mathbf{p_{y}}+M\widetilde{\omega}\mathbf{x}+iM\omega \mathbf{y}\big)\varPsi_{10}&=&0, \label{spin1-2} \\
Mc^{2}\varPsi_{3}+E\varPsi_{9}&=&0, \label{spin1-3} \\
Mc^{2}\varPsi_{4}+ic\big(\mathbf{p_{y}}+M\widetilde{\omega}\mathbf{x}-iM\omega \mathbf{y}\big)\varPsi_{9}&=&0, \label{spin1-4} \\
Mc^{2}\varPsi_{5}-ic\big(\mathbf{p_{x}}-M\widetilde{\omega}\mathbf{y}-iM\omega \mathbf{x}\big)\varPsi_{9}&=&0, \label{spin1-5} \\
ic\big(\mathbf{p_{y}}+M\widetilde{\omega}\mathbf{x}-iM\omega \mathbf{y}\big) \varPsi_{7}-ic\big(\mathbf{p_{x}}-M\widetilde{\omega}\mathbf{y}-iM\omega \mathbf{x}\big)\varPsi_{8}-Mc^2\varPsi_{6}&=&0, \label{spin1-6} \\
-Mc^{2}\varPsi_{7}-E\varPsi_{1}+ic\big(\mathbf{p_{y}}+M\widetilde{\omega}\mathbf{x}+iM\omega \mathbf{y}\big)\varPsi_{6}&=&0, \label{spin1-7} \\
Mc^{2}\varPsi_{8}+E\varPsi_{2}+ic\big(\mathbf{p_{x}}-M\widetilde{\omega}\mathbf{y}+iM\omega \mathbf{x}\big)\varPsi_{6}&=&0, \label{spin1-8} \\
-ic\big(\mathbf{p_{y}}+M\widetilde{\omega}\mathbf{x}+iM\omega \mathbf{y}\big) \varPsi_{4}+ic\big(\mathbf{p_{x}}-M\widetilde{\omega}\mathbf{y}+iM\omega \mathbf{x}\big)\varPsi_{5}-E\varPsi_{3}-Mc^{2}\varPsi_{9}&=&0, \label{spin1-9} \\
ic\big(\mathbf{p_{x}}-M\widetilde{\omega}\mathbf{y}-iM\omega \mathbf{x}\big) \varPsi_{1}+ic\big(\mathbf{p_{y}}+M\widetilde{\omega}\mathbf{x}-iM\omega \mathbf{y}\big)\varPsi_{2}+Mc^{2}\varPsi_{10}&=&0. \label{spin1-10}
\end{eqnarray}
To our knowledge, the analytic solutions to these coupled equation remain unknown. Instead, some authors employed Eqs. (\ref{spin1-3}), (\ref{spin1-4}), (\ref{spin1-5}) and (\ref{spin1-9}) to build a correlation among the third, fourth, fifth and ninth components \cite{31, Wu_et_al_2017}. Moreover, without giving any arguments  nor physical justifications they made an ansatz and assumed that $\varPsi_{1}=\varPsi_{2}=0$, thus, $\varPsi_{6}=\varPsi_{7}=\varPsi_{8}=\varPsi_{10}=0$.  With this ansatz, six of the ten equations vanish identically and the equation for the ninth component appears as
\begin{eqnarray}
&&\bigg[c^2\Big(\mathbf{p_{x}}-M\widetilde{\omega}\mathbf{y}+iM\omega \mathbf{x}\Big)\Big(\mathbf{p_{x}}-M\widetilde{\omega}\mathbf{y}-iM\omega \mathbf{x}\Big)\nonumber \\&+&
c^2\Big(\mathbf{p_{y}}+M\widetilde{\omega}\mathbf{x}+iM\omega \mathbf{y}\Big)\Big(\mathbf{p_{y}}+M\widetilde{\omega}\mathbf{x}-iM\omega \mathbf{y}\Big) +\Big(M^2c^{4}-E^{2}\Big)\bigg]\varPsi_{9}=0. \label{S1main}
\end{eqnarray}%
Note that Eq. (\ref{S1main}) is equivalent to Eq. (\ref{S0main}). Therefore, the energy spectrum functions in the spin-one and spin-zero cases become identical to each other. However, this can be accepted only as a very particular solution.

Instead of following that ansatz, we examine the spin-one case in another formalism as given in \cite{Falek_et_al_2010, Falek_et_al_2017, Falek_et_al_2019}. We rewrite the ten component spinor by using three vectors with three components and a scalar as follows:
\begin{eqnarray}
\varPsi^T\equiv\Big(\begin{array}{cccc}
\mathbf{B},& \mathbf{C},& \mathbf{A}, & \varphi
\end{array}\Big).
\end{eqnarray}%
Here the vectors are defined with the following components
\begin{eqnarray}
\mathbf{A}&\equiv&
\big(\begin{array}{ccc}
\varPsi_{1}, & \varPsi_{2}, & \varPsi_{3}
\end{array}\big), \\
\mathbf{B}&\equiv&
\big(\begin{array}{ccc}
\varPsi_{4}, & \varPsi_{5}, &\varPsi_{6}
\end{array}\big), \\
\mathbf{C}&\equiv&\big(\begin{array}{ccc}
\varPsi_{7}, & \varPsi_{8}, & \varPsi_{9}
\end{array}\big), \\
\varphi &\equiv& \varPsi_{10}.
\end{eqnarray}%
Then, we define two new vectors with three components
\begin{eqnarray}
\mathbf{P^+}&\equiv&
\big(\begin{array}{ccc}
Q_{1}, & Q_{2}, & 0
\end{array}\big), \\
\mathbf{P^-}&\equiv&
\big(\begin{array}{ccc}
Q_{3}, & Q_{4}, &0
\end{array}\big),
\end{eqnarray}%
where
\begin{eqnarray}
  Q_1 &\equiv& ic\big(\mathbf{p_{x}}-M\widetilde{\omega}\mathbf{y}+iM\omega \mathbf{x}\big), \\
  Q_2 &\equiv& ic\big(\mathbf{p_{y}}+M\widetilde{\omega}\mathbf{x}+iM\omega \mathbf{y}\big), \\
  Q_3 &\equiv& ic\big(\mathbf{p_{x}}-M\widetilde{\omega}\mathbf{y}-iM\omega \mathbf{x}\big), \\
  Q_4 &\equiv& ic\big(\mathbf{p_{y}}+M\widetilde{\omega}\mathbf{x}-iM\omega \mathbf{y}\big).
\end{eqnarray}
With the help of these definitions, we reduce the set of equations to
\begin{eqnarray}
  Mc^2\mathbf{B}+E\mathbf{A}+\mathbf{P^{+}}\varphi &=& 0, \\
  Mc^2\mathbf{C}+\mathbf{P^{-}}\times \mathbf{A} &=& 0, \\
  Mc^2\mathbf{A}+E\mathbf{B}-\mathbf{P^{+}}\times \mathbf{C} &=& 0, \\
  \mathbf{P^{-}}\cdot \mathbf{B} +Mc^2\varphi &=& 0.
\end{eqnarray}
We decouple the equations by eliminating $\mathbf{B}$,  $\mathbf{C}$, and $\varphi$ in terms of $\mathbf{A}$. We get
\begin{eqnarray}
  &&(E^2-M^2c^4)\mathbf{A} +\Big[(\mathbf{P^{-}} \cdot \mathbf{P^{+}})\mathbf{A} -(\mathbf{P^{+}}\times\mathbf{P^{-}})\times\mathbf{A} \Big]+\frac{1}{M^2c^4}\mathbf{P^{+}}\Big[\mathbf{P^{-}}\cdot \Big(\mathbf{P^{+}}\times \big( \mathbf{P^{-}}\times \mathbf{A}\big)\Big)\Big]=0.
\end{eqnarray}
Next, we evaluate $(\mathbf{P^{-}} \cdot \mathbf{P^{+}})\mathbf{A}$ and $(\mathbf{P^{+}}\times\mathbf{P^{-}})\times\mathbf{A}$, as done in \cite{Falek_et_al_2019}. We find
\begin{eqnarray}
 (\mathbf{P^{-}} \cdot \mathbf{P^{+}})\mathbf{A} &=&-c^2\bigg[ \big(\mathbf{p_{x}}\mathbf{p_{x}}+\mathbf{p_{y}}\mathbf{p_{y}}\big)+2 M\widetilde{\omega}\mathbf{L_z}
      +M^2\big({\widetilde{\omega}}^2+\omega^2\big)\big(\mathbf{x}\mathbf{x}+\mathbf{y}\mathbf{y}\big)-2 M\hbar\omega \bigg]\mathbf{A},\,\,\,\,\,\,\,\,\,\,\,\,\\
  (\mathbf{P^{+}}\times\mathbf{P^{-}})\times\mathbf{A}&=& -c^2\bigg[2M^2\omega\widetilde{\omega} \big(\mathbf{x}\mathbf{x}+\mathbf{y}\mathbf{y}\big)+2M \omega \mathbf{L_{z}} -2M\hbar \widetilde{\omega}\bigg]\frac{\mathbf{s_z}}{\hbar}\mathbf{A},
\end{eqnarray}
where $[\mathbf{x},\mathbf{p_{x}}]=i\hbar$ and $\mathbf{L_z}=\mathbf{x}\mathbf{p_{y}}-\mathbf{y}\mathbf{p_{x}}$. Similar to \cite{Falek_et_al_2019}, by substituting the terms we get
\begin{eqnarray}
  &&(E^2-M^2c^4)\mathbf{A} -c^2\bigg[ \big(\mathbf{p_{x}}\mathbf{p_{x}}+\mathbf{p_{y}}\mathbf{p_{y}}\big)+2 M\mathbf{L_z}\bigg( \widetilde{\omega}-\omega\frac{\mathbf{s_z}}{\hbar}\bigg)
      +M^2\big(\mathbf{x}\mathbf{x}+\mathbf{y}\mathbf{y}\big)\bigg({\widetilde{\omega}}^2+\omega^22\omega\widetilde{\omega}
      \frac{\mathbf{s_z}}{\hbar}\bigg) \nonumber \\
      &-&2 M\hbar \bigg(\omega-\widetilde{\omega}\frac{\mathbf{s_z}}{\hbar}\bigg) \bigg]\mathbf{A}-
      + \frac{1}{M^2c^4}\mathbf{P^{+}}\Big[\mathbf{P^{-}}\cdot \Big(\mathbf{P^{+}}\times \big( \mathbf{P^{-}}\times \mathbf{A}\big)\Big)\Big]=0.
\end{eqnarray}
To the best of our knowledge, there is no known exact solution to this equation \cite{Falek_et_al_2019}.
\subsubsection{The non-relativistic limit}
We go to the non-relativistic limit
\begin{eqnarray}
  \frac{(E^2-M^2c^4)}{2Mc^2}\mathbf{A} &-&\bigg[\frac{\big(\mathbf{p_{x}}\mathbf{p_{x}}+\mathbf{p_{y}}\mathbf{p_{y}}\big)}{2M}+\mathbf{L_z}\bigg( \widetilde{\omega}-\omega\frac{\mathbf{s_z}}{\hbar}\bigg)+\frac{M}{2}\big(\mathbf{x}\mathbf{x}+\mathbf{y}\mathbf{y}\big)\bigg({\widetilde{\omega}}^2+\omega^2-
  2\omega\widetilde{\omega}\frac{\mathbf{s_z}}{\hbar}  \bigg) \nonumber \\
      &-&\hbar \bigg(\omega-\widetilde{\omega}\frac{\mathbf{s_z}}{\hbar}\bigg) \bigg]\mathbf{A} +\frac{1}{2(Mc^2)^3}\mathbf{P^{+}}\Big[\mathbf{P^{-}}\cdot \Big(\mathbf{P^{+}}\times \big( \mathbf{P^{-}}\times \mathbf{A}\big)\Big)\Big]=0.
\end{eqnarray}
Then, we ignore the last term since it is in order of $(Mc^2)^{-3}$ and use the non-zero eigenvalues of $\mathbf{s_z}$, $(m_s=\hbar, 0, -\hbar)$ \cite{Falek_et_al_2019}
\begin{eqnarray}
  \bigg[\frac{\mathbf{p^2}}{2M}+\frac{M\big({\widetilde{\omega}}^2+\omega^2\pm4\omega\widetilde{\omega}  \big)}{2}\mathbf{r}^2+
  \mathbf{L_z}\big( \widetilde{\omega}\pm2\omega\big)-\hbar \big(\omega\pm2\widetilde{\omega}\big) \bigg]\mathbf{A} &=&E^{NRL}_{n, m}\mathbf{A}. \label{DKP_ML_bir}
\end{eqnarray}
We find the non relativistic energy eigenvalue function as follows \cite{Valentim_et_al_2019}:
\begin{eqnarray}
 E^{NRL}_{n, m}&=& \hbar\Big[\sqrt{{\widetilde{\omega}}^2+\omega^2\pm4\omega\widetilde{\omega}}\big(2n+{|m|}+1\big)+m\big( \widetilde{\omega}\pm2\omega\big)- \big(\omega\pm2\widetilde{\omega}\big)\Big]. \label{ENRL-spin1}
\end{eqnarray}
We conclude that the energy spectrum function in the spin-one case differs from the spin-zero case in the non-relativistic limit.

\section{DKP oscillator in the ML quantum mechanics}  \label{MLQMDKP}

In the ML formalism, we consider the modified Heisenberg algebra generated by the coordinate, $\mathbf{\hat{X}_{i}}$, and the momentum, $\mathbf{\hat{P}_{i}}$,  operators which satisfy the algebra given in Eqs. (\ref{C1}), (\ref{C2}), and (\ref{C3}).  We employ the Heisenberg algebra representation on the momentum space where the momentum and position operators act on a momentum space wave function as given \cite{2}
\begin{eqnarray}
\mathbf{\hat{P}_{i}} \varPsi(p) &=& p_{i} \varPsi(p), \label{ML_momentum_op}\\
\mathbf{\hat{X}_{i}} \varPsi(p) &=& i\hbar\Big(1+\beta p^{2}\Big)\frac{\partial}{\partial p_i} \varPsi(p).\label{ML_position_op}
\end{eqnarray}
Note that the ML parameter has the inverse square of the momentum unit. Before we proceed through the following section, we would like to remind the definition of the inner product as well \cite{2}.
\begin{eqnarray}
\big(\varPsi,\varPhi\big)=\int \frac{d^3p}{1+\beta p^2}\varPsi^*(p)\varPhi(p).
\end{eqnarray}

\subsection{Case of the spin-zero particle} \label{ML_DKP-0}
In this subsection, we investigate the DKP oscillator that is examined in Sec. \ref{DKP-0} in the ML formalism. Since we assume that the oscillator is under the effect of an equivalent external magnetic field, we obtain the same five coupled equations that are given in Eqs. (\ref{spin-0_eq_1}), (\ref{spin-0_eq_2}), (\ref{spin-0_eq_3}), (\ref{spin-0_eq_4}), and (\ref{spin-0_eq_5}), thus, Eq. (\ref{S0main}). We employ the momentum and position operators, which are defined by the ML formalism with Eq. (\ref{ML_momentum_op}) and Eq. (\ref{ML_position_op}), in the Eq. (\ref{S0main}). In the polar coordinates, we find a differential equation for the first component of the wave function as follows
\begin{eqnarray}
&& \Bigg\{p^2-2M\hbar\Big(1+\beta p^2\Big)\Bigg[\bigg(\omega  +i \widetilde{\omega} \frac{ \partial }{\partial \theta}\bigg)+\beta M\hbar\bigg( \Omega^2  p \frac{\partial}{\partial p}-2i\widetilde{\omega}\omega \frac{\partial}{\partial \theta}\bigg)\Bigg] \nonumber \\
&-&  M^2\hbar ^2\Omega^2  \big(1+\beta p^2\big)^2 \bigg(\frac{\partial^{2}}{\partial p^{2}}+\frac{1}{p}\frac{\partial}{\partial p}+ \frac{1}{p^{2}} \frac{\partial^{2}}{\partial\theta^{2}}\bigg)-\varsigma \Bigg\}\varPsi_{1}=0. \label{DKPspin0-ML-1}
\end{eqnarray}%
Note that, when the ML parameter is taken to be zero, Eq. (\ref{DKPspin0-ML-1}) reduces to Eq. (\ref{DKPspin0-1}). Next, we decouple the wave function to spatial and angular parts as given in Eq. (\ref{Spat_angular}), and use it in Eq. (\ref{DKPspin0-ML-1}). We find
\begin{eqnarray}
&& \Bigg\{p^2-2M\hbar\Big(1+\beta p^2\Big)\bigg[\big(\omega  - m \widetilde{\omega}\big)+\beta M\hbar\bigg( \Omega^2  p \frac{d}{dp}+2m\widetilde{\omega}\omega \bigg)\bigg] \nonumber \\
&-&  M^2\hbar ^2\Omega^2  \big(1+\beta p^2\big)^2 \bigg(\frac{d^{2}}{d p^{2}}+\frac{1}{p}\frac{d}{d p}- \frac{m^2}{p^{2}} \bigg)-\varsigma \Bigg\}f(p)=0. \label{DKPspin0-ML-2}
\end{eqnarray}%
Then, we follow the paper of Jana \emph{et al.} \cite{Jana_et_al_2009}. There, they showed that a second order differential equation in the form
\begin{eqnarray}
\bigg[-a(p)\frac{d^2}{dp^2}+b(p)\frac{d}{dp}+c(p)\bigg]\phi(p)&=&\varsigma \phi(p), \label{Jana}
\end{eqnarray}
is transformed to a Schr\"odinger-type differential equation by employing
\begin{eqnarray}
\zeta(p)&\equiv& \frac{\frac{da(p)}{dp}+2b(p)}{4a(p)}, \label{J1}\\
\rho(p)&\equiv& \exp{\int \zeta(p)dp},  \label{J2}\\
\phi(p)&\equiv& \rho(p) \varphi(p). \label{J3}
\end{eqnarray}
We match Eq. (\ref{Jana}) with Eq. (\ref{DKPspin0-ML-2}) to determine $a(p)$, $b(p)$, and $c(p)$ functions. We find
\begin{eqnarray}
a(p)&=&M^2\hbar ^2\Omega^2  \big(1+\beta p^2\big)^2, \\
b(p)&=& -2\beta M^2\hbar ^2\Omega^2 p \big(1+\beta p^2\big) - \frac{M^2\hbar ^2\Omega^2  \big(1+\beta p^2\big)^2}{p}, \\
c(p)&=&p^2-2M\hbar\big(1+\beta p^2\big)\Big(\big(\omega  - m \widetilde{\omega}\big)+2\beta M \hbar m \widetilde{\omega} \omega\Big)+\frac{M^2\hbar ^2\Omega^2  \big(1+\beta p^2\big)^2m^2}{p^2}.
\end{eqnarray}
We calculate $\zeta(p)$, then, $\rho(p)$, and obtain
\begin{eqnarray}
\phi(p)&=& \frac{1}{\sqrt{p}} \varphi(p), \label{ph}
\end{eqnarray}
which converts Eq. (\ref{Jana}) to
\begin{eqnarray}
\bigg[-a(p)\frac{d^2}{dp^2}+\bigg(\frac{a(p)}{p}+b(p)\bigg)\frac{d}{dp}+\bigg(c(p)-\frac{3a(p)}{4p^2}-\frac{b(p)}{2p}\bigg)\bigg]\varphi(p)&=&\varsigma \varphi(p). \label{Jana1}
\end{eqnarray}
Next, we use a variable change
\begin{eqnarray}
q&\equiv&\int \frac{dp}{\sqrt{a(p)}}, \label{var1}
\end{eqnarray}
that transforms the $p$ coordinate to $q$ coordinate via
\begin{eqnarray}
p&=& \frac{1}{\sqrt{\beta}} \tan \Big(M\hbar \Omega \sqrt{\beta} q \Big), \label{var2}
\end{eqnarray}
in Eq. (\ref{Jana1}). After a straightforward calculation we obtain
\begin{eqnarray}
-\frac{d^2\chi(q)}{dq^2}+\beta M^2\hbar ^2\Omega^2\Bigg[\frac{\xi_1(\xi_1-1)}{\sin^2 \Big(M\hbar \Omega \sqrt{\beta} q \Big)}+\frac{\xi_2(\xi_2-1)}{\cos^2 \Big(M\hbar \Omega \sqrt{\beta} q \Big)}\Bigg]\chi(q)&=&\sigma\chi(q),
\end{eqnarray}
where
\begin{eqnarray}
\sigma&\equiv& \varsigma+\frac{1}{\beta},\\
\xi_1(\xi_1-1)&\equiv& \Big(m^2-\frac{1}{4}\Big), \label{xi1}\\
\xi_2(\xi_2-1)&\equiv& \frac{1}{\beta^2 M^2\hbar ^2\Omega^2 } - \frac{2\big(\omega  - m \widetilde{\omega}\big)}{\beta M\hbar \Omega^2}-\frac{4 \widetilde{\omega} \omega m}{\Omega^2}  +  \Big(m^2+\frac{3}{4}\Big). \label{xi2}
\end{eqnarray}
We introduce a new coordinate transformation of the form
\begin{eqnarray}
z&\equiv&\sin^2 \Big(M\hbar \Omega \sqrt{\beta} q \Big),
\end{eqnarray}
and we get
\begin{eqnarray}
z(1-z)\frac{d^2 u(z)}{dz^2}+\Big(\frac{1}{2}-z\Big)\frac{d u(z)}{dz}+\frac{1}{4}\bigg[\frac{\sigma}{\beta M^2\hbar ^2\Omega^2} -\frac{\xi_1(\xi_1-1)}{z}-\frac{\xi_2(\xi_2-1)}{1-z}\bigg]u(z)&=&0.\,\,\,
\end{eqnarray}
We propose the general solution as an Ansatz
\begin{eqnarray}
u(z)&\equiv& z^{{\frac{|\xi_1|}{2}}} (1-z)^{\frac{|\xi_2|}{2}}v(z),
\end{eqnarray}
and we obtain the hypergeometric equation in the following form of
\begin{eqnarray}
&&z(1-z)\frac{d^2 v(z)}{dz^2}+\bigg[\Big({|\xi_1|}+\frac{1}{2}\Big)-z\big(1+{|\xi_1|}+{|\xi_2|}\big)\bigg]\frac{d v(z)}{dz}-\frac{1}{4}\bigg[\big({|\xi_1|}+{|\xi_2|}\big)-\frac{\sigma}{\beta M^2\hbar ^2\Omega^2}\bigg]v(z)=0.\nonumber \\
\end{eqnarray}
The general solution of the hypergeometric function is given as \cite{Abramowitz_et_al_Book}
\begin{eqnarray}
&&v(z)=C_{1}\,_{2}F_1\Bigg(\frac{1}{2} \bigg({|\xi_1|}+{|\xi_2|}+\frac{1}{M\hbar\Omega} \sqrt{\frac{\sigma}{\beta}}  \bigg), \frac{1}{2} \bigg({|\xi_1|}+{|\xi_2|}-\frac{1}{M\hbar\Omega} \sqrt{\frac{\sigma}{\beta}}  \bigg), {|\xi_1|}+\frac{1}{2},z\Bigg) \nonumber \\
&+& C_{2} z^{\frac{1}{2}-{|\xi_1|}} \,_{2}F_1\Bigg(\frac{1}{2} \bigg(1-{|\xi_1|}+{|\xi_2|}+\frac{1}{M\hbar\Omega} \sqrt{\frac{\sigma}{\beta}}  \bigg),\frac{1}{2} \bigg(1-{|\xi_1|}+{|\xi_2|}-\frac{1}{M\hbar\Omega} \sqrt{\frac{\sigma}{\beta}}  \bigg),\frac{3}{2}-{|\xi_1|},z \Bigg),
\end{eqnarray}
where $C_1$ and $C_2$ are normalization constants. Note that, Eqs. (\ref{xi1}) and (\ref{xi2}) are quadratic equations, thus they have two roots. In both, we take the larger root from the quadratic formula. {To avoid a singularity, we set $C_2=0$}. Then, we employ the quantization condition
\begin{eqnarray}
\frac{1}{2} \bigg({|\xi_1|}+{|\xi_2|}+\frac{1}{M\hbar\Omega} \sqrt{\frac{\sigma}{\beta}}  \bigg)&=&-n,
\end{eqnarray}
where $n$ is an integer. After simple calculus, we obtain the energy function
\begin{eqnarray}
E_{n,m}=\pm Mc^{2}\left[1+\frac{2\hbar\left(m\widetilde{\omega}-\omega\right)}{Mc^{2}}+\frac{\hbar\Omega\left(4n+2\left({|m|}+1\right)\right)}
{Mc^{2}}{|\sqrt{\alpha_m}|}+\frac{{|\Lambda_{n,m}|}}{M^{2}c^{4}}\right]^{\frac{1}{2}}, \label{Energy_ML_spin0}
\end{eqnarray}
where
\begin{eqnarray}
\alpha_m&\equiv&\left[1+2\beta M\hbar\left(m\widetilde{\omega}-\omega\right)+\beta^{2}M^{2}\hbar^{2}\left(\Omega^{2}\left(m^{2}+1\right)-4m\widetilde{\omega}\omega\right)\right], \label{Energy_ML_spin0a} \\
\Lambda_{n,m}&\equiv& \beta\left(Mc\hbar\Omega\right)^{2}\Big[\big(2n+m+1\big)^2+m^2+1-\frac{4m\widetilde{\omega}\omega}{\Omega^{2}}\Big].\label{Energy_ML_spin0b}
\end{eqnarray}
We would like to emphasize that in the limit of  $\beta=0$, $\alpha$ goes to $1$ while $\Lambda_{n,m}$ vanishes. Therefore, Eq. (\ref{Energy_ML_spin0}) turns to  Eq. (\ref{Energy_spin0}), which means we end with the habitual quantum mechanical result. Moreover, in another limit, where the external magnetic field does not exist, the cyclotron frequency vanishes and the energy eigenvalue function becomes identical to the one given in Eq. (34) in  \cite{Wu_et_al_2017}.

\subsubsection{The non-relativistic limit}
In this limit, the energy spectrum function reduces to
\begin{eqnarray}
 E^{NRL}_{n, m}&=& \hbar\Big[\sqrt{{\widetilde{\omega}}^2+\omega^2}\big(2n+{|m|}+1\big){|\sqrt{\alpha_m}|}+\big(m\widetilde{\omega}-\omega\big)\Big]
 +{\big|\Lambda_{n,m}^{NRL}\big|},
\end{eqnarray}
where
\begin{eqnarray}
\Lambda_{n,m}^{NRL}=\frac{\beta M}{2}\left(\hbar\Omega\right)^{2}\Big[\big(2n+m+1\big)^2+m^2+1-\frac{4m\widetilde{\omega}\omega}{\Omega^{2}}\Big].
\end{eqnarray}

\subsection{Case of the spin-one particle} \label{ML_DKP-1}
In this subsection, we examine the DKP oscillator for a spin-one particle in the ML formalism in the non-relativistic limit. We employ Eqs.(\ref{ML_momentum_op}) and (\ref{ML_position_op}) in Eq.( \ref{DKP_ML_bir}) with the polar coordinates' definitions given with Eqs. (\ref{xop}), (\ref{yop}), (\ref{pox}) and (\ref{poy}). Then, we allocate the wave function into the spatial and angular functions. For the spatial part, we get
\begin{eqnarray}
  &&\Bigg[p^2-M^2\big({\widetilde{\omega}}^2+\omega^2\pm4\omega\widetilde{\omega}\big)\bigg[\hbar^2\big(1+\beta p^2\big)^2\bigg(\frac{d^2}{dp^2}+\frac{1}{p}\frac{d}{dp}-\frac{m^2}{p^2}\bigg)+2\beta\hbar^2 p \big(1+\beta p^2\big)\frac{d}{dp}\bigg] \nonumber \\
  &+&2mM\hbar({\widetilde{\omega}}\pm 2\omega)\big(1+\beta p^2\big)-2M \hbar ({\omega \pm 2 \widetilde{\omega}})-2M E^{NRL} \Bigg]g(p)=0. \label{ML_DKP_bir-1}
\end{eqnarray}
Similarly to the spin-zero case, we follow the paper of Jana \emph{et al.} \cite{Jana_et_al_2009}. We define ${\widetilde{\Omega}}^2\equiv \big({\widetilde{\omega}}^2+\omega^2\pm4\omega\widetilde{\omega}\big)$ and compare Eq. (\ref{ML_DKP_bir-1}) with Eq. (\ref{Jana}). We obtain
\begin{eqnarray}
  a(p) &=& M^2 \hbar^2 {\widetilde{\Omega}}^2\big(1+\beta p^2\big)^2, \\
  b(p) &=& -M^2 \hbar^2 {\widetilde{\Omega}}^2\frac{\big(1+\beta p^2\big)^2}{p}  -2\beta M^2 \hbar^2{\widetilde{\Omega}}^2  p\big(1+\beta p^2\big),\\
  c(p) &=& p^2+ m^2 M^2 \hbar^2 {\widetilde{\Omega}}^2\frac{\big(1+\beta p^2\big)^2}{p^2}+ 2mM\hbar({\widetilde{\omega}}\pm 2\omega)\big(1+\beta p^2\big)-2M \hbar ({\omega \pm 2 \widetilde{\omega}}).
\end{eqnarray}
Then, we use Eqs. (\ref{J1}), (\ref{J2}) and (\ref{J3}) to calculate Eq. (\ref{ph}). We change the variables as given in Eqs. (\ref{var1}) and (\ref{var2}) and get
\begin{eqnarray}
  \Bigg[-\frac{d^2}{dq^2}+ \beta M^2\hbar ^2{\widetilde{\Omega}}^2\Bigg(\frac{\widetilde{\xi}_1(\widetilde{\xi}_1-1)}{\sin^2 \Big(M\hbar {\widetilde{\Omega}} \sqrt{\beta} q \Big)}+\frac{\widetilde{\xi}_2(\widetilde{\xi}_2-1)}{\cos^2 \Big(M\hbar {\widetilde{\Omega}} \sqrt{\beta} q \Big)}\Bigg)\Bigg]\chi(q) &=&\widetilde{\sigma}\chi(q),
\end{eqnarray}
where
\begin{eqnarray}
\widetilde{\sigma}&\equiv& 2M\big(E^{NRL}+\hbar(\omega\pm 2\widetilde{\omega})\big)+\frac{1}{\beta}, \label{tildesigma} \\
\widetilde{\xi}_1(\widetilde{\xi}_1-1)&\equiv& \Big(m^2-\frac{1}{4}\Big), \label{ML1_xi1}\\
\widetilde{\xi}_2(\widetilde{\xi}_2-1)&\equiv& \frac{1}{\beta^2 M^2\hbar ^2\widetilde{\Omega}^2 }+ \frac{2 m\big(\widetilde{\omega}\pm 2\omega \big)}{\beta M\hbar \widetilde{\Omega}^2}  +  \bigg(m^2+\frac{3}{4}\bigg). \label{ML1_xi2}
\end{eqnarray}
The roots of Eqs. (\ref{ML1_xi1}) and  (\ref{ML1_xi2}) are
\begin{eqnarray}
  \widetilde{\xi}_1 &=& \pm m+ \frac{1}{2},  \label{rootmlx1} \\
  \widetilde{\xi}_2  &=& \pm \sqrt{\widetilde{\kappa}_m}+ \frac{1}{2}, \label{rootmlx2}
\end{eqnarray}
where
\begin{eqnarray}
 \widetilde{\kappa}_m &\equiv&  \frac{1+2\beta M m\hbar\big(\widetilde{\omega}\pm 2\omega \big)}{\beta^2 M^2\hbar ^2\widetilde{\Omega}^2 } +  \big(m^2+1 \big).
\end{eqnarray}
We prefer to choose the positive root between them. Then, we introduce the transformation $z\equiv \sin^2 \Big(M\hbar \widetilde{\Omega} \sqrt{\beta} q \Big)$. We find
\begin{eqnarray}
z(1-z)\frac{d^2 u(z)}{dz^2}+\Big(\frac{1}{2}-z\Big)\frac{d u(z)}{dz}+\frac{1}{4}\bigg[\frac{\widetilde{\sigma}}{\beta M^2\hbar ^2\widetilde{\Omega}^2} -\frac{\widetilde{\xi}_1(\widetilde{\xi}_1-1)}{z}-\frac{\widetilde{\xi}_2(\widetilde{\xi}_2-1)}{1-z}\bigg]u(z)&=&0.\,\,\,\,\,\,
\end{eqnarray}
For the general solution we make an Ansatz
\begin{eqnarray}
u(z)&\equiv& z^{{\frac{|\widetilde{\xi}_1|}{2}}} (1-z)^{{\frac{|\widetilde{\xi}_2|}{2}}}v(z).
\end{eqnarray}
We obtain
\begin{eqnarray}
&&z(1-z)\frac{d^2 v(z)}{dz^2}+\bigg[\Big({|\widetilde{\xi}_1|}+\frac{1}{2}\Big)-z\big(1+{|\widetilde{\xi}_1|}+{|\widetilde{\xi}_2|}\big)\bigg]\frac{d v(z)}{dz}\nonumber \\
&-&\frac{1}{4}\bigg[\big({|\widetilde{\xi}_1|}+{|\widetilde{\xi}_2|}\big)-\frac{\widetilde{\sigma}}{\beta M^2\hbar ^2\widetilde{\Omega}^2}\bigg]v(z)=0.
\end{eqnarray}
This type of equation admits solutions in terms of hypergeometric functions \cite{Abramowitz_et_al_Book}.
\begin{eqnarray}
&&v(z)=D_{1}\,_{2}F_1\Bigg(\frac{1}{2} \bigg({|\widetilde{\xi}_1|}+{|\widetilde{\xi}_2|}+\frac{1}{M\hbar\widetilde{\Omega}} \sqrt{\frac{\widetilde{\sigma}}{\beta}}  \bigg), \frac{1}{2} \bigg({|\widetilde{\xi}_1|}+{|\widetilde{\xi}_2|}-\frac{1}{M\hbar\widetilde{\Omega}} \sqrt{\frac{\widetilde{\sigma}}{\beta}}  \bigg), {|\widetilde{\xi}_1|}+\frac{1}{2},z\Bigg), \\
&+& D_{2} z^{\frac{1}{2}-{|\widetilde{\xi}_1|}} \,_{2}F_1\Bigg(\frac{1}{2} \bigg(1-{|\widetilde{\xi}_1|}+{|\widetilde{\xi}_2|}+\frac{1}{M\hbar\widetilde{\Omega}} \sqrt{\frac{\widetilde{\sigma}}{\beta}}  \bigg),\frac{1}{2} \bigg(1-{|\widetilde{\xi}_1|}+\widetilde{\xi}_2-\frac{1}{M\hbar\widetilde{\Omega}} \sqrt{\frac{\widetilde{\sigma}}{\beta}}  \bigg),\frac{3}{2}-{|\widetilde{\xi}_1|},z \Bigg).\nonumber
\end{eqnarray}
Here $D_1$ and $D_2$ are the normalization constants. {For the regularity of the wave function at $z=0$, we choose $D_2=0$}. The quantization condition gives
\begin{eqnarray}
{|\widetilde{\xi}_1|}+{|\widetilde{\xi}_2|}+\frac{1}{M\hbar\widetilde{\Omega}} \sqrt{\frac{\widetilde{\sigma}}{\beta}} &=&-2n,
\end{eqnarray}
where $n$ is an integer. Finally, we employ Eqs. (\ref{tildesigma}), (\ref{rootmlx1}) and (\ref{rootmlx2}) in the quantization condition and derive the energy eigenvalue function as
\begin{eqnarray}
 E^{NRL}_{n, m}&=& \hbar\Big[\sqrt{{\widetilde{\omega}}^2+\omega^2\pm4\omega\widetilde{\omega}}\big(2n+{|m|}+1\big){|\sqrt{\widetilde{\alpha}_m}|}+m\big( \widetilde{\omega}\pm2\omega\big)- \big(\omega\pm2\widetilde{\omega}\big)\Big]+{\big|\widetilde{\Lambda}_{n,m}^{NRL}\big|}, \label{MLENRLspin-1}
\end{eqnarray}
where
\begin{eqnarray}
\widetilde{\alpha}_m &=& 1+2\beta M m \hbar\left(\widetilde{\omega}\pm2\omega\right)+ \beta^{2}M^{2}\hbar^{2}{\widetilde{\Omega}}^{2}\left(m^2+1\right), \\
\widetilde{\Lambda}_{n,m}^{NRL}  &=& \frac{\beta M}{2}\left(\hbar\widetilde{\Omega}\right)^{2}\Big[\big(2n+m+1\big)^2+m^2+1\Big].
\end{eqnarray}
Note that, when $\beta=0$, the result becomes identical to the OQM result, which is given in Eq. (\ref{ENRL-spin1}).

\section{Thermal Properties} \label{TP}
In this section, we investigate the statistical properties of the two-dimensional DKP oscillator that is under the effect of an external magnetic field in the presence of an ML.  Although we derived spectrum functions for spin-zero and spin-one cases in the previous sections, for the sake of the length of the manuscript, we will merely investigate the thermal properties of the spin-zero particle ensemble at high-temperatures. We assume that the DKP oscillator is in an equilibrium state at a finite temperature, $T$. We employ the definition of the partition function of the canonical ensemble
\begin{eqnarray}
Z&\equiv&\sum_{j=0}^{\infty}{\exp{\bigg( -\frac{E_j}{k_B T} \bigg)}}. \label{Partitionfunction}
%Z&\equiv&\sum_{n=0}^{\infty}{\sum_{m=-n}^n} \exp{\bigg( -\frac{E_{n,m}}{k_B T} \bigg)}. \label{Partitionfunction}
\end{eqnarray}
Here, $k_B$ is the  Boltzmann constant. We establish the thermodynamic ensemble by considering the states with positive eigenvalues out of all values in the energy spectrum, because we do not accept the negative energy states as physical states since they are not bounded from below. Thus, we guarantee that the stability of the ensemble \cite{Pacheco_et_al_2014}. We take into account the energy spectrum that is derived in Eq. (\ref{Energy_ML_spin0}) with the correlated expressions given in Eqs. (\ref{Energy_ML_spin0a}) and (\ref{Energy_ML_spin0b}).  We use
\begin{eqnarray}
 y_{n,m}^2&\equiv&1+\frac{2\hbar\big( m \widetilde{\omega} - \omega \big)}{Mc^2} + \frac{\hbar \Omega \Big(4n+2\big({\big|m|}+1\big)\Big)}{Mc^2}{\big| \sqrt{\alpha_m}\big|} + {\Big|\frac{\Lambda_{n,m}}{M^2c^4}\Big|},
 \\
 \gamma&\equiv&\frac{Mc^2}{k_B T},
\end{eqnarray}
to express the partition function as
\begin{eqnarray}
Z&=&\sum_{n=0}^{\infty} F(n),
\end{eqnarray}
where
\begin{eqnarray}
  F(n) &\equiv& \sum_{m=-n}^{n} \exp \Big(-\gamma y_{n,m}\Big).
\end{eqnarray}
We employ the Euler-Mclaurin summation formula to compute the partition function as done in \cite{Falek_et_al_2019, Falek_et_al_2017, Nouicer_2006}.
\begin{eqnarray}
\sum_{n=0}^{\infty} F\left(n\right)=\frac{1}{2}F\left(0\right)+ \int_{0}^{\infty}dnF(n)-\sum_{\rho=1}^{\infty}\frac{B_{2\rho}}{\left(2\rho\right)!} F^{\left(2\rho-1\right)}(0), \label{EMLform}
\end{eqnarray}%
Here, $B_{2\rho}$ represents the Bernoulli numbers while $F^{\left(2\rho-1\right)}$ denotes the order of the derivative. First, we execute the integral term
\begin{eqnarray}
\int_{0}^{\infty}dnF(n)&=&\frac{Mc^{2}}{2\hbar\Omega}\int_{y_{0,0}}^{+\infty}y_{n,m}dy_{n,m}\bigg[\frac{1}{1-\beta\left(Mc\right)^{2}(1-y_{n,m}^2)}\bigg]^{\frac{1}{2}}\exp{\big(-\gamma y_{n,m}\big)},
\end{eqnarray}
where
\begin{eqnarray}
  y_{0,0}&=&\sqrt{1-\frac{2\hbar\omega}{Mc^2} + \frac{2\hbar \omega \sqrt{1+\frac{\widetilde{\omega}^2}{\omega^2} }}{Mc^2} \Big|1-\beta M \hbar \omega \Big|\sqrt{1+\bigg(\frac{\beta M\hbar \widetilde{\omega}}{1-\beta M \hbar \omega}\bigg)^2} +  \frac{2\beta M^2c^2\hbar^2 (\omega^2+\widetilde{\omega}^2)}{(Mc^2)^2}} . \nonumber \\
\end{eqnarray}
We evaluate the integral in the high temperature region just after we expand the square root to its Taylor series. We obtain
\begin{eqnarray}
 \int_{0}^{\infty}dnF(n)&=& \frac{Mc^{2}}{2\hbar\Omega\sqrt{1-\beta\left(Mc\right)^{2}}}\bigg(\frac{k_B T}{Mc^2}\bigg)^2\sum_{\rho=0}\Gamma\left(2\rho+2\right) \frac{\left(2\rho\right)!}{\left(2^{\rho}\rho!\right)^{2}}\left(\frac{\beta\left(Mc\right)^{2}}{\beta\left(Mc\right)^{2}-1}\right)^{\rho} \label{sum}
 \bigg(\frac{k_B T}{Mc^2}\bigg)^{2\rho}.\nonumber \\
\end{eqnarray}
Then, we focus on the first and third terms in Eq. (\ref{EMLform}). They are constituted from $F(n)\Big|_{n=0}$, $ \frac{dF(n)}{dn}\Big|_{n=0}$, $\frac{d^3F(n)}{dn^3}\Big|_{n=0}$,  $\cdots $ functions, which depend on the reciprocal temperature. Therefore at higher temperatures,  their contributions become negligible and we ignore them \cite{Falek_et_al_2019, Falek_et_al_2017, Nouicer_2006}.

\begin{figure}[!htb]
\minipage{0.80\textwidth}
  \includegraphics[width=0.90\linewidth]{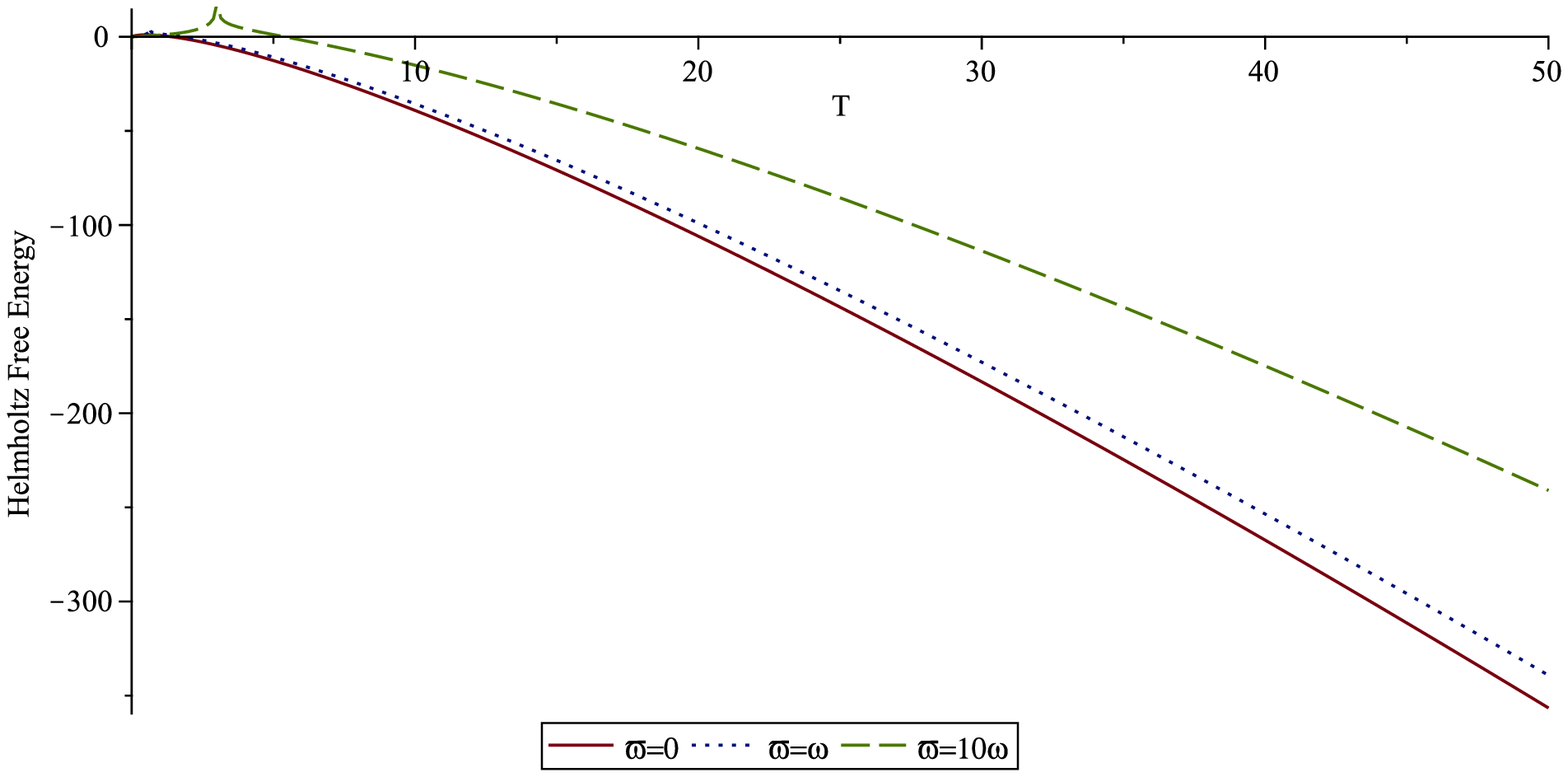}
  \subcaption{$\beta=0$}
\endminipage\hfill\\
\minipage{0.80\textwidth}
  \includegraphics[width=0.90\linewidth]{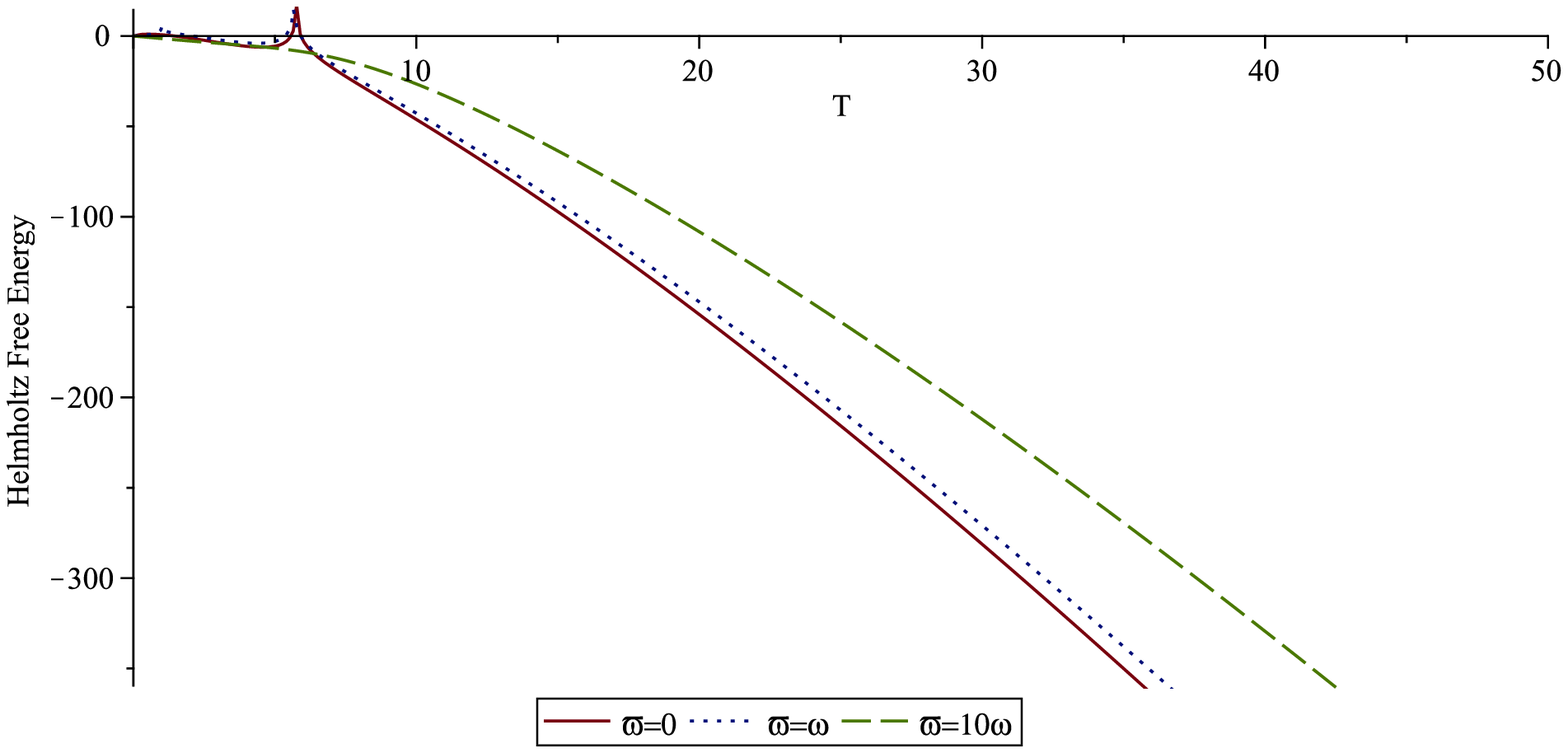}
  \subcaption{$\beta=0.005$}
\endminipage\hfill \\
\minipage{0.80\textwidth}%
  \includegraphics[width=0.90\linewidth]{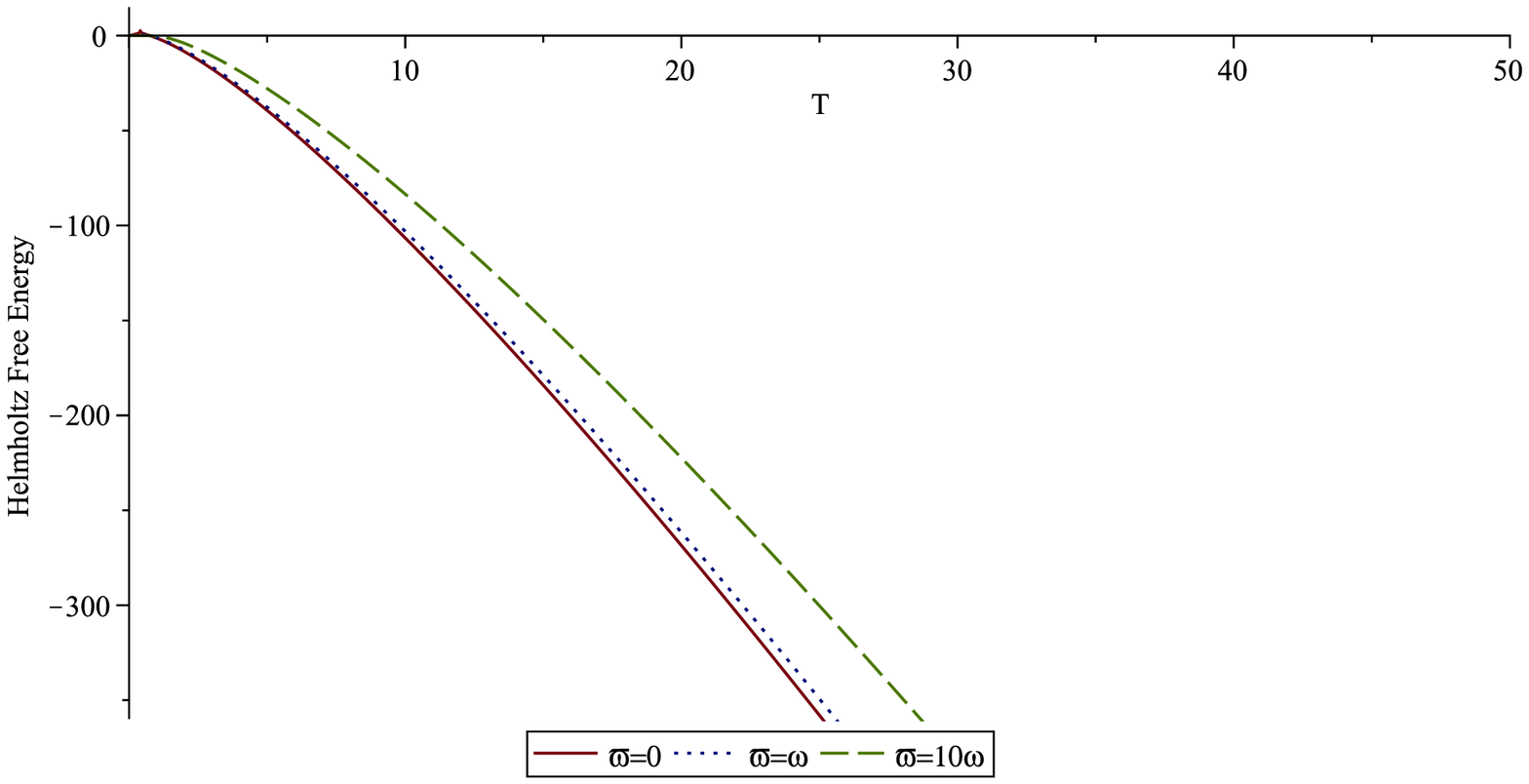}
  \subcaption{$\beta=0.5$}
\endminipage
\caption{The behavior of the Helmholtz free energy function of the DKP oscillator versus the reduced temperature for different values of the minimal length parameter.} \label{HFE}
\end{figure}

\begin{figure}[!htb]
\minipage{0.80\textwidth}
  \includegraphics[width=0.90\linewidth]{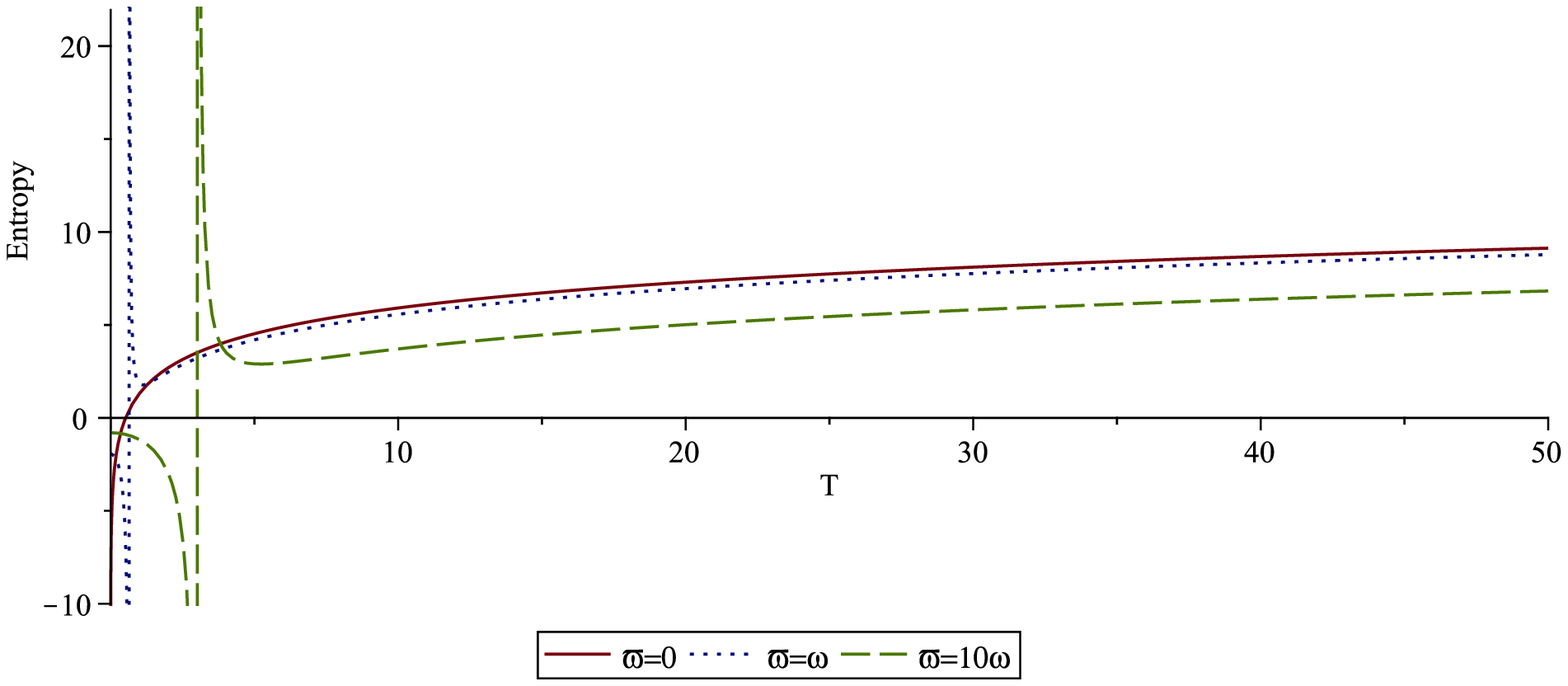}
  \subcaption{$\beta=0$}
\endminipage\hfill\\
\minipage{0.80\textwidth}
  \includegraphics[width=0.90\linewidth]{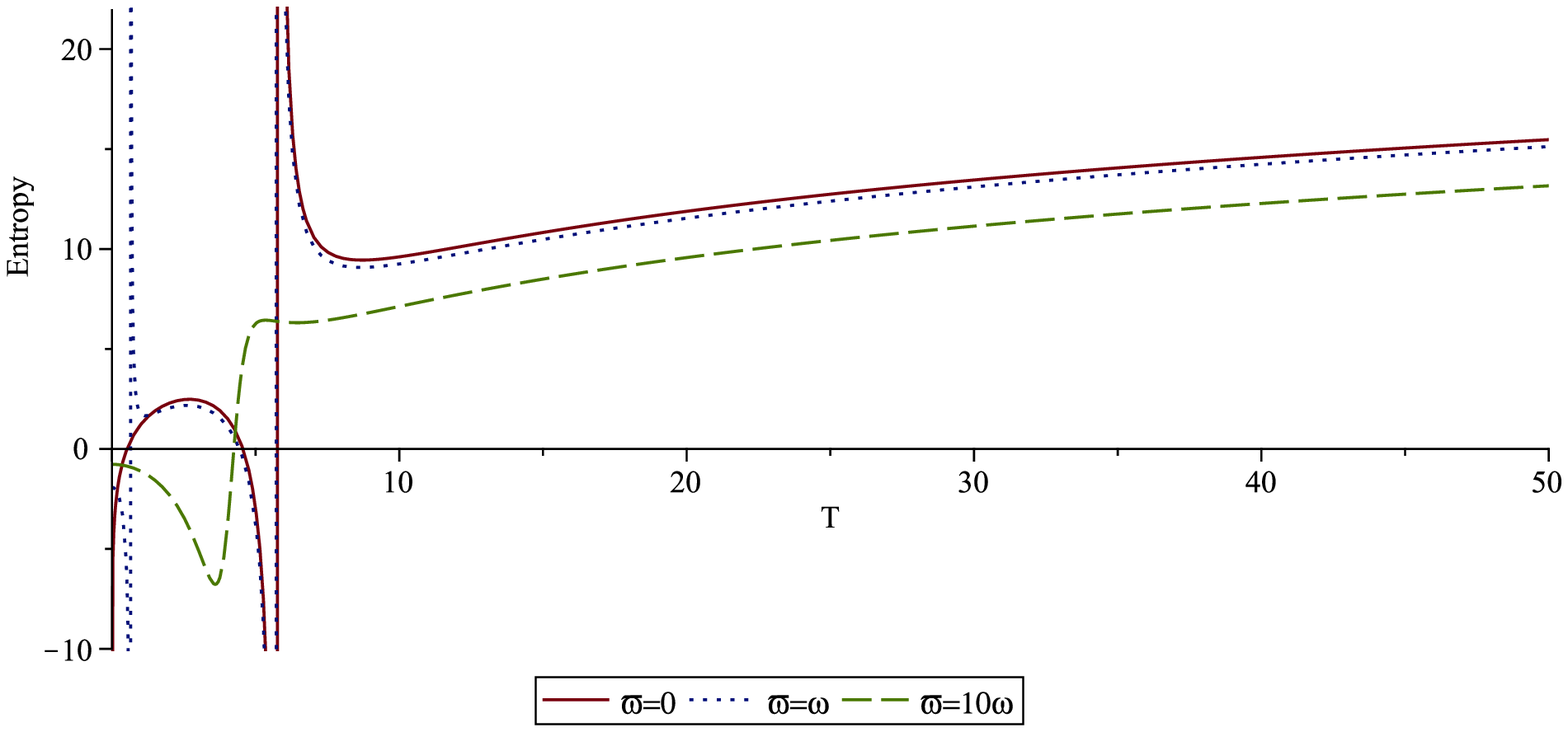}
  \subcaption{$\beta=0.005$}
\endminipage\hfill \\
\minipage{0.80\textwidth}%
  \includegraphics[width=0.90\linewidth]{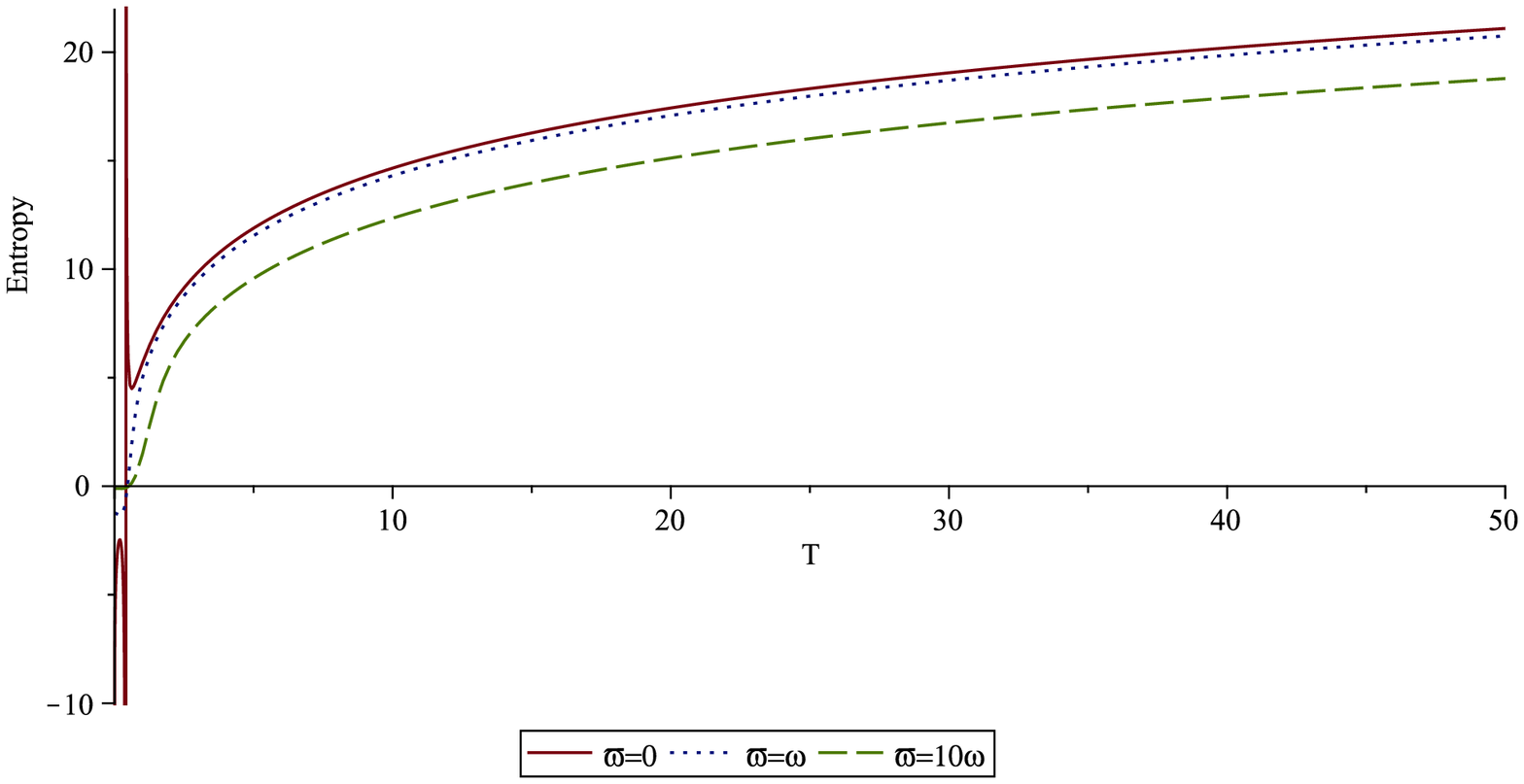}
  \subcaption{$\beta=0.5$}
\endminipage
\caption{The behavior of the entropy function of the DKP oscillator versus the reduced temperature for different values of the minimal length parameter.} \label{Ent}
\end{figure}

\begin{figure}[!htb]
\minipage{0.80\textwidth}
  \includegraphics[width=0.80\linewidth]{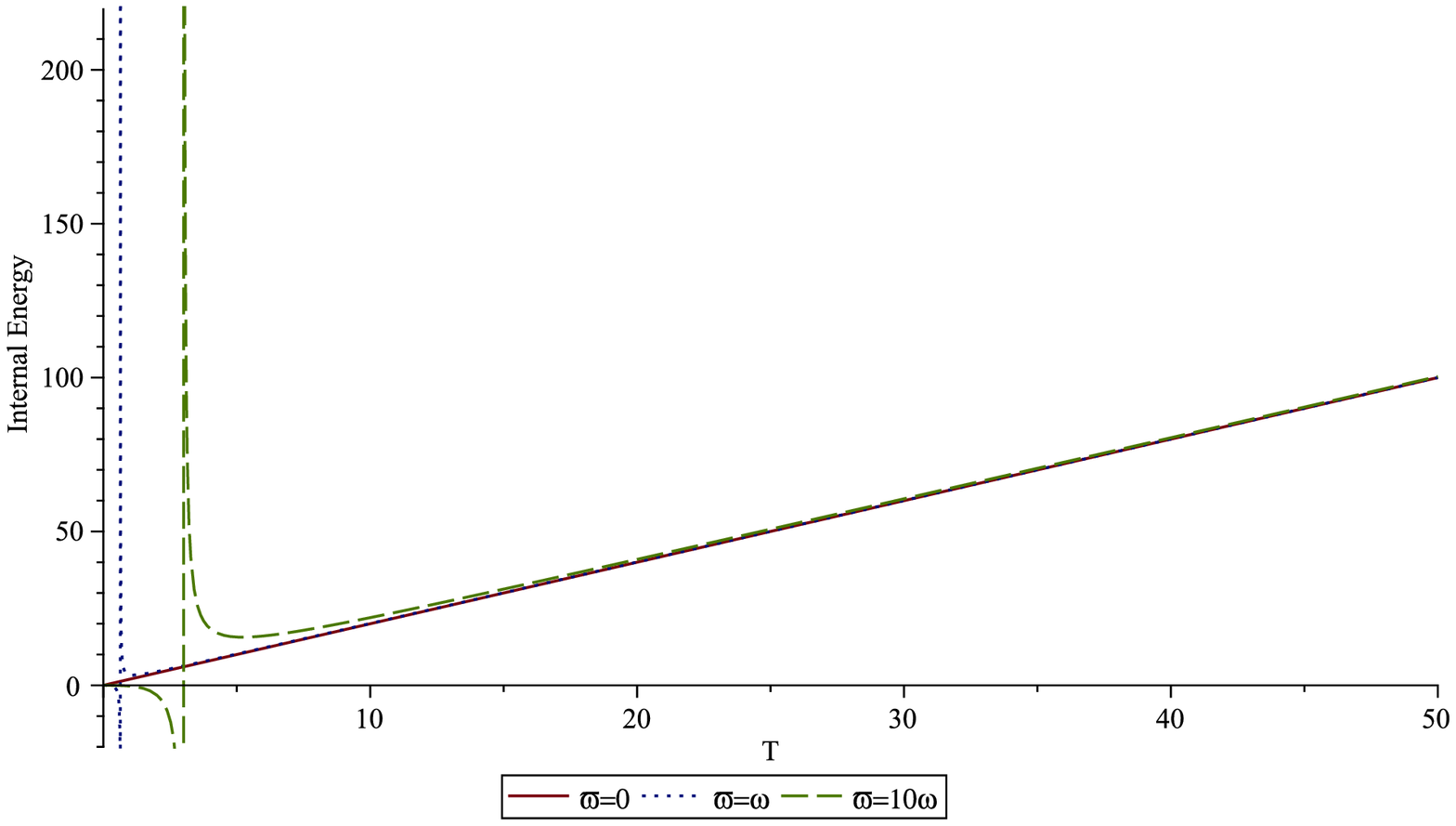} \label{b0000U}
  \subcaption{$\beta=0$}
\endminipage\hfill\\
\minipage{0.80\textwidth}
  \includegraphics[width=0.80\linewidth]{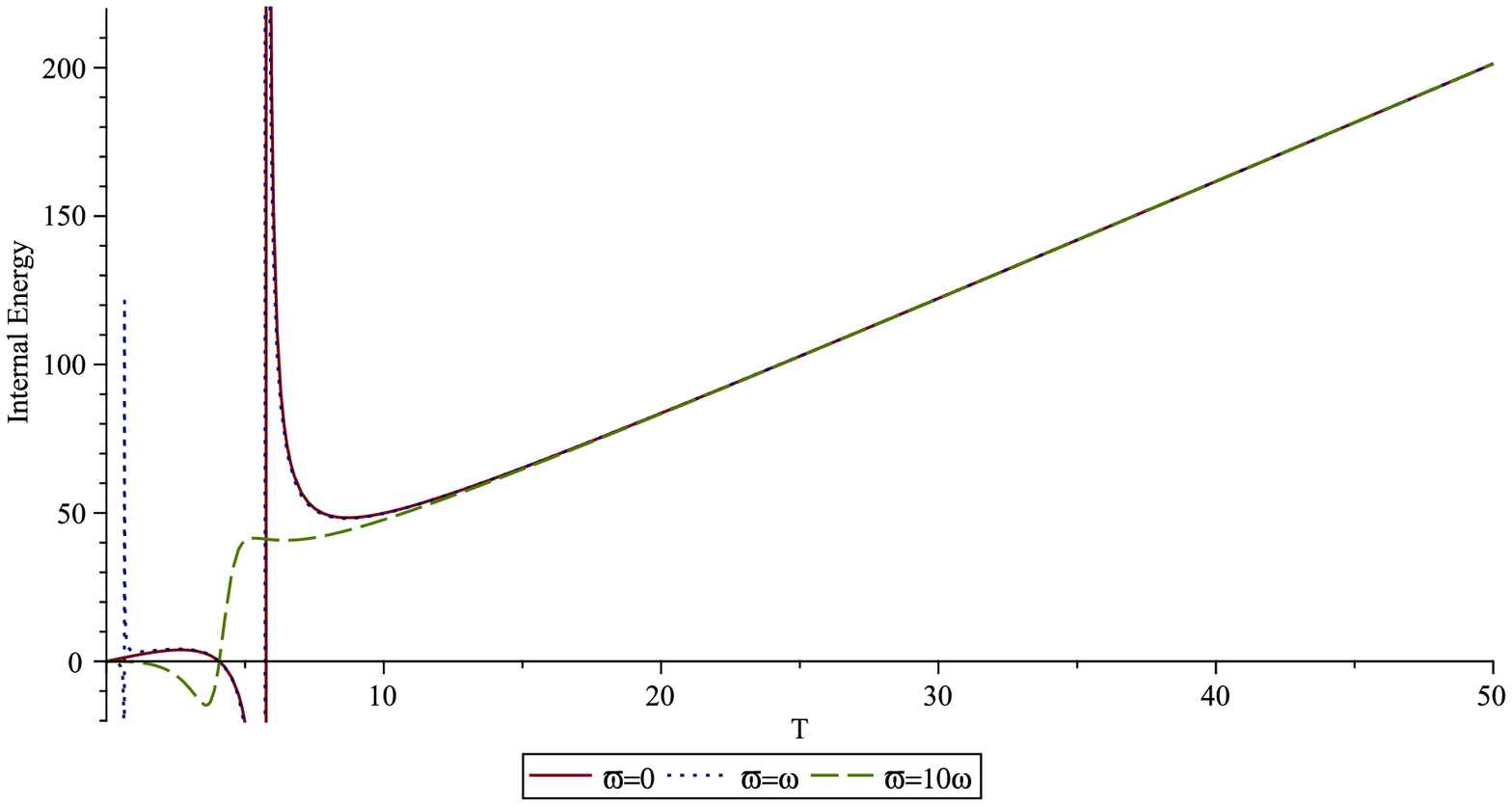}
  \subcaption{$\beta=0.005$}
\endminipage\hfill \\
\minipage{0.80\textwidth}%
  \includegraphics[width=0.80\linewidth]{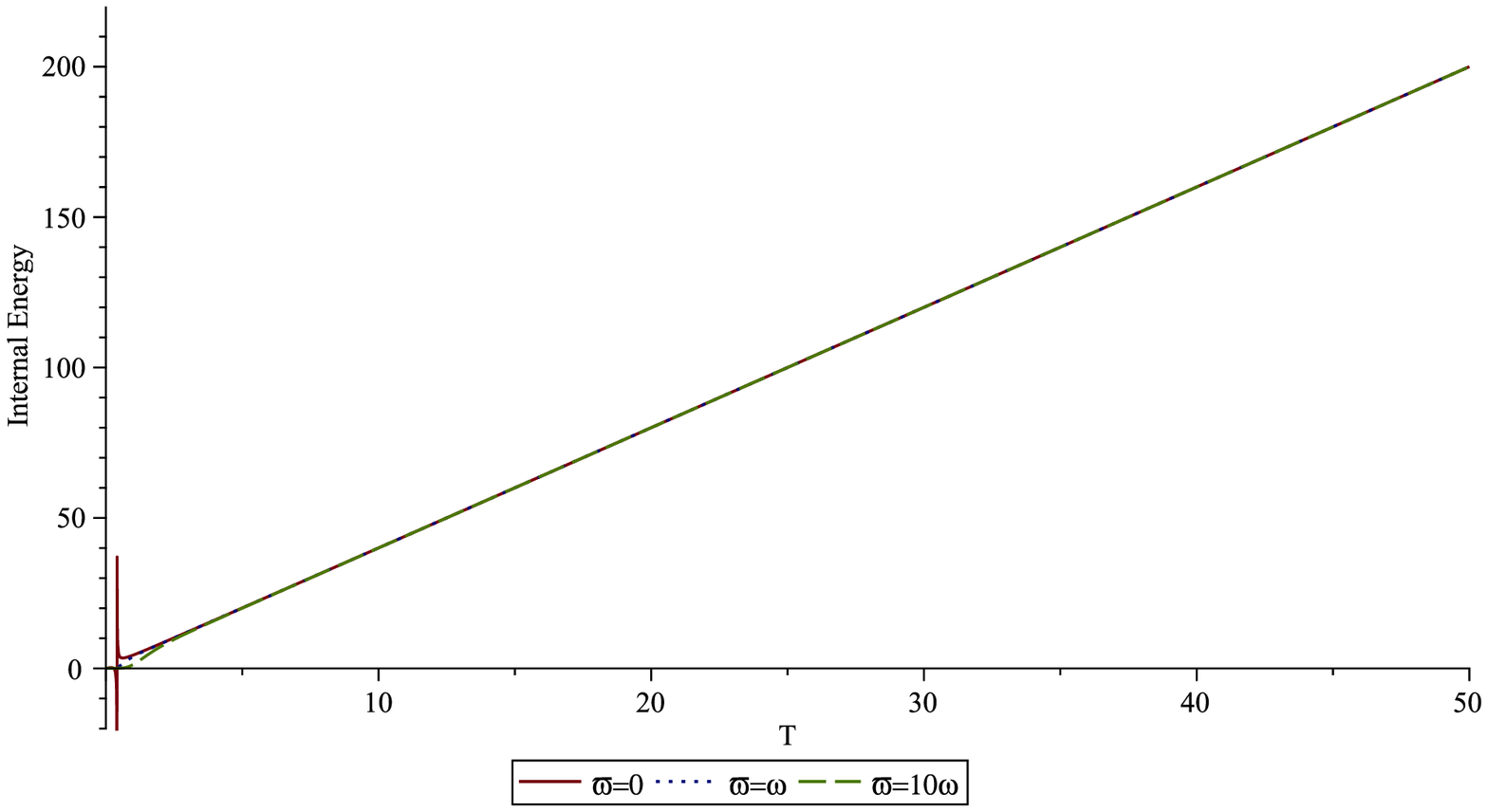}
  \subcaption{$\beta=0.5$}
\endminipage
\caption{The behavior of the internal energy function of the DKP oscillator versus the reduced temperature for different values of the minimal length parameter.} \label{Int}
\end{figure}

\begin{figure}[!htb]
\minipage{0.80\textwidth}
  \includegraphics[width=0.90\linewidth]{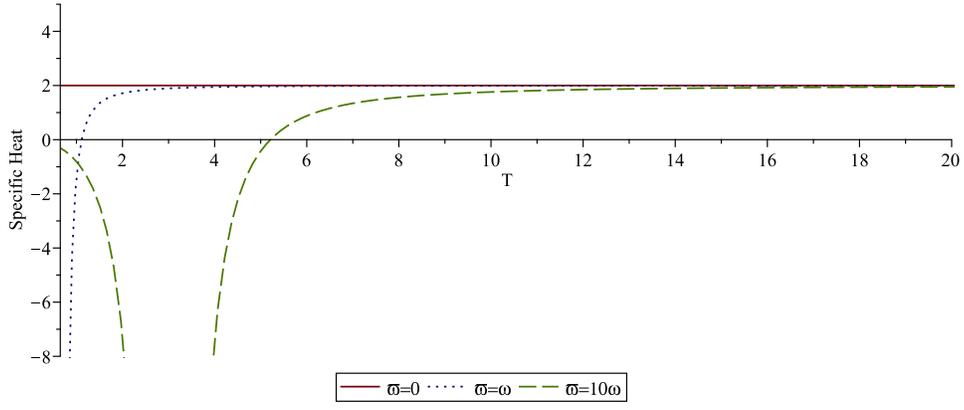}
  \subcaption{$\beta=0$}
\endminipage\hfill\\
\minipage{0.80\textwidth}
  \includegraphics[width=0.90\linewidth]{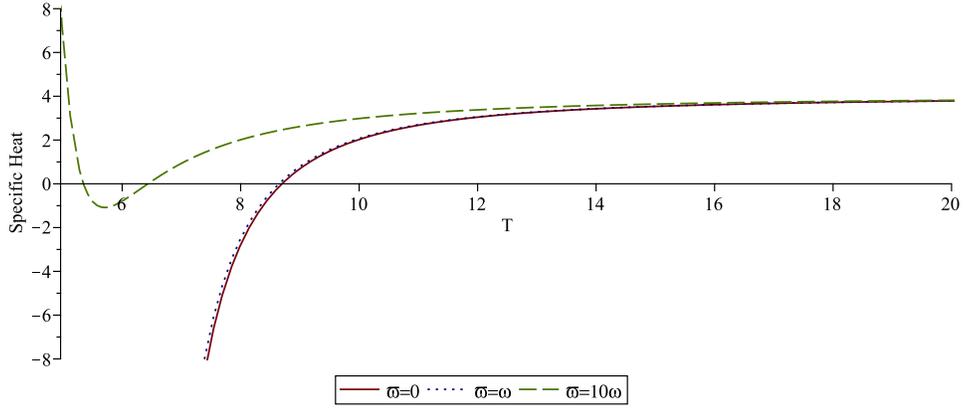}
  \subcaption{$\beta=0.005$}
\endminipage\hfill \\
\minipage{0.80\textwidth}%
  \includegraphics[width=0.90\linewidth]{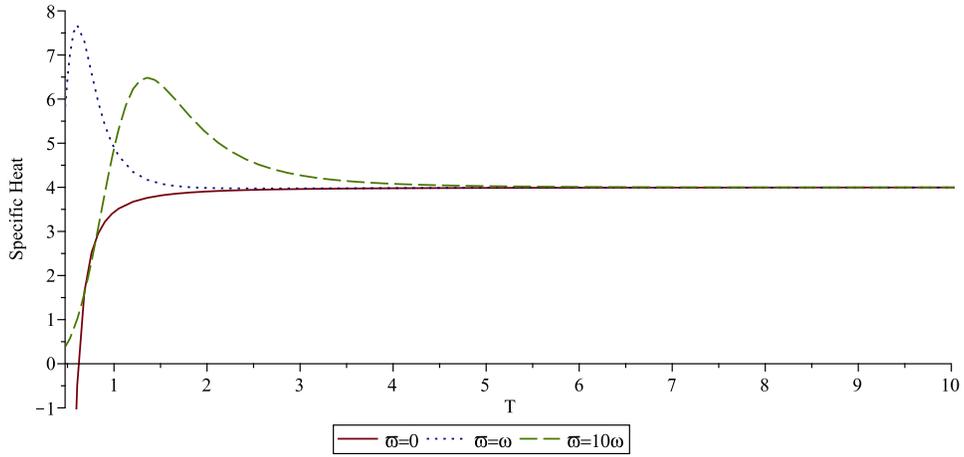}
  \subcaption{$\beta=0.5$}
\endminipage
\caption{The behavior of the specific heat function of the DKP oscillator versus the reduced temperature for different values of the minimal length parameter.} \label{SpH}
\end{figure}

As the final step, we expand the summation given in Eq. (\ref{sum}) and  keep the terms only up to the first order of the ML parameter. We obtain the partition function as
\begin{eqnarray}
Z&=&\frac{Mc^{2}}{2\hbar\Omega\sqrt{1-\beta\left(Mc\right)^{2}}}\Bigg[\left(\frac{k_{B}T}{Mc^{2}}\right)^{2}+6\left(\frac{\beta\left(Mc\right)^{2}}{\beta\left(Mc\right)^{2}-1}\right)
\left(\frac{k_{B}T}{Mc^{2}}\right)^{4}\Bigg] \nonumber\\
&-&\Bigg[\frac{\sqrt{1-\beta\left(Mc\right)^{2}\left(1-y_{0,0}^{2}\right)}-1}{2\beta M\hbar\Omega}\Bigg]. \label{MLPF}
\end{eqnarray}%
Next, we employ the following definitions
\begin{eqnarray}
F   &\equiv&  - k_{B}Tln\left(Z\right), \\
S   &\equiv&  - \frac{\partial F}{\partial T}, \\
U   &\equiv&    k_BT^2 \frac{\partial}{\partial T} \ln Z, \\
C_v &\equiv&    \frac{\partial U}{\partial T},
\end{eqnarray}%
to derive the Helmholtz free energy, entropy, internal energy and specific heat functions, respectively. In high-temperature region, we find the Helmholtz free energy function as
\begin{eqnarray}
F&=&-k_B T \ln \Bigg\{\frac{Mc^{2}}{2\hbar\Omega\sqrt{1-\beta\left(Mc\right)^{2}}}\Bigg[\left(\frac{k_{B}T}{Mc^{2}}\right)^{2}+6\left(\frac{\beta\left(Mc\right)^{2}}
{\beta\left(Mc\right)^{2}-1}\right)\left(\frac{k_{B}T}{Mc^{2}}\right)^{4}\Bigg]\nonumber \\
 &&-\Bigg[\frac{\sqrt{1-\beta\left(Mc\right)^{2}\left(1-y_{0,0}^{2}\right)}-1}{2\beta M\hbar\Omega}\Bigg] \Bigg\}, \label{F}
\end{eqnarray}%
the entropy function as
\begin{eqnarray}
&&S = k_B  \ln \Bigg\{\frac{Mc^{2}}{2\hbar\Omega\sqrt{1-\beta\left(Mc\right)^{2}}}\Bigg[\left(\frac{k_{B}T}{Mc^{2}}\right)^{2}+6\left(\frac{\beta\left(Mc\right)^{2}}{\beta\left(Mc\right)^{2}-1}\right)
\left(\frac{k_{B}T}{Mc^{2}}\right)^{4}\Bigg]\nonumber \\
&-&\Bigg[\frac{\sqrt{1-\beta\left(Mc\right)^{2}\left(1-y_{0,0}^{2}\right)}-1}{2\beta M\hbar\Omega}\Bigg] \Bigg\} \nonumber \\
&+& \frac{k_B\frac{Mc^{2}}{\hbar\Omega\sqrt{1-\beta\left(Mc\right)^{2}}}\Bigg[\left(\frac{k_{B}T}{Mc^{2}}\right)^{2}+12\left(\frac{\beta\left(Mc\right)^{2}}{\beta\left(Mc\right)^{2}-1}\right)
\left(\frac{k_{B}T}{Mc^{2}}\right)^{4}\Bigg]}{\frac{Mc^{2}}{2\hbar\Omega\sqrt{1-\beta\left(Mc\right)^{2}}}\Bigg[\left(\frac{k_{B}T}{Mc^{2}}\right)^{2}+6\left(\frac{\beta\left(Mc\right)^{2}}{\beta\left(Mc\right)^{2}-1}\right)
\left(\frac{k_{B}T}{Mc^{2}}\right)^{4}\Bigg]-\Bigg[\frac{\sqrt{1-\beta\left(Mc\right)^{2}\left(1-y_{0,0}^{2}\right)}-1}{2\beta M\hbar\Omega}\Bigg]},
 \label{Entropy}
\end{eqnarray}%
the internal energy function as
\begin{eqnarray}
U&=&\frac{2k_B T \Bigg[\left(\frac{k_{B}T}{Mc^{2}}\right)^{2}+12\left(\frac{\beta\left(Mc\right)^{2}}{\beta\left(Mc\right)^{2}-1}\right)
\left(\frac{k_{B}T}{Mc^{2}}\right)^{4}\Bigg]}{\Bigg[\left(\frac{k_{B}T}{Mc^{2}}\right)^{2}+6\left(\frac{\beta\left(Mc\right)^{2}}{\beta\left(Mc\right)^{2}-1}\right)
\left(\frac{k_{B}T}{Mc^{2}}\right)^{4}\Bigg]-\Bigg[\frac{\big(\sqrt{1-\beta\left(Mc\right)^{2}}\big)\big(\sqrt{1-\beta\left(Mc\right)^{2}\left(1-y_{0,0}^{2}\right)}-1\big)}{\beta (Mc)^{2}}\Bigg]}, \label{InternalEnergy}
\end{eqnarray}%
and the specific heat function as
\begin{eqnarray}
&&C_v = \frac{6k_B  \Bigg[\left(\frac{k_{B}T}{Mc^{2}}\right)^{2}+20\left(\frac{\beta\left(Mc\right)^{2}}{\beta\left(Mc\right)^{2}-1}\right)
\left(\frac{k_{B}T}{Mc^{2}}\right)^{4}\Bigg]}{\Bigg[\left(\frac{k_{B}T}{Mc^{2}}\right)^{2}+6\left(\frac{\beta\left(Mc\right)^{2}}{\beta\left(Mc\right)^{2}-1}\right)
\left(\frac{k_{B}T}{Mc^{2}}\right)^{4}\Bigg]-\Bigg[\frac{\big(\sqrt{1-\beta\left(Mc\right)^{2}}\big)\big(\sqrt{1-\beta\left(Mc\right)^{2}\left(1-y_{0,0}^{2}\right)}-1\big)}{\beta (Mc)^{2}}\Bigg]} \nonumber \\
&-& \frac{4k_B  \Bigg[\left(\frac{k_{B}T}{Mc^{2}}\right)^{2}+12\left(\frac{\beta\left(Mc\right)^{2}}{\beta\left(Mc\right)^{2}-1}\right)
\left(\frac{k_{B}T}{Mc^{2}}\right)^{4}\Bigg]^2}{\left[\Bigg(\left(\frac{k_{B}T}{Mc^{2}}\right)^{2}+6\left(\frac{\beta\left(Mc\right)^{2}}{\beta\left(Mc\right)^{2}-1}\right)
\left(\frac{k_{B}T}{Mc^{2}}\right)^{4}\Bigg)-\Bigg(\frac{\big(\sqrt{1-\beta\left(Mc\right)^{2}}\big)\big(\sqrt{1-\beta\left(Mc\right)^{2}\left(1-y_{0,0}^{2}\right)}-1\big)}{\beta (Mc)^{2}}\Bigg)\right]^2}. \label{Cv}
\end{eqnarray}
Note that, in the OQM limit where $\beta=0$, these functions reduce to
\begin{eqnarray}
Z&=&\frac{Mc^2}{2\hbar \Omega}\bigg(\frac{k_B T}{Mc^2}\bigg)^2+\frac{1}{2}\bigg(\frac{\omega}{\Omega}-1\bigg),\label{OQMLZ} \\
F&=&-k_BT \ln  \Bigg[\frac{Mc^2}{2\hbar \Omega}\bigg(\frac{k_B T}{Mc^2}\bigg)^2+\frac{1}{2}\bigg(\frac{\omega}{\Omega}-1\bigg)\Bigg], \label{OQMLHFE}\\
S&=&k_B  \left\{ \ln\left[\frac{Mc^2}{2\hbar \Omega}\bigg(\frac{k_B T}{Mc^2}\bigg)^2+\frac{1}{2}\bigg(\frac{\omega}{\Omega}-1\bigg) \right] + 2 \left[1- \frac{\bigg(\frac{\omega}{\Omega}-1\bigg)}{\bigg[\frac{Mc^2}{\hbar \Omega}\bigg(\frac{k_B T}{Mc^2}\bigg)^2+\bigg(\frac{\omega}{\Omega}-1\bigg)\bigg]}\right]\right\},
 \label{OQMEntropy} \\
U&=&2k_B T  \left[1-\frac{\Big(\frac{\omega}{\Omega}-1\Big)}{\frac{Mc^2}{\hbar \Omega}\Big(\frac{k_B T}{Mc^2}\Big)^2+\Big(\frac{\omega}{\Omega}-1\Big)} \right],\\
C_v&=&2k_B\left\{1+\left[\frac{\Big(\frac{\omega}{\Omega}-1\Big)}{{\frac{Mc^2}{\hbar \Omega}\Big(\frac{k_B T}{Mc^2}\Big)^2+\Big(\frac{\omega}{\Omega}-1\Big)}}\right]-2\left[\frac{\Big(\frac{\omega}{\Omega}-1\Big)}{{\frac{Mc^2}{\hbar \Omega}\Big(\frac{k_B T}{Mc^2}\Big)^2+\Big(\frac{\omega}{\Omega}-1\Big)}}\right]^2\right\}.
\end{eqnarray}
Furthermore, in the absence of the external magnetic field in the OQM limit we obtain the well-known thermodynamic functions of a scalar bosonic oscillator \cite{Falek_et_al_2019}.
\begin{eqnarray}
Z&=&\frac{Mc^2}{2\hbar \omega}\bigg(\frac{k_B T}{Mc^2}\bigg)^2, \\
F&=&-k_BT \ln  \Bigg[\frac{Mc^2}{2\hbar \Omega}\bigg(\frac{k_B T}{Mc^2}\bigg)^2\Bigg], \\
S&=& k_B  \left\{ \ln\left[\frac{Mc^2}{2\hbar \Omega}\bigg(\frac{k_B T}{Mc^2}\bigg)^2 \right] + 2 \right\}, \\
U&=& 2k_B T, \label{OQMNMFU}\\
C_v&=&2k_B.
\end{eqnarray}
Then, we plot the thermodynamic functions versus the temperature for different values of the ML parameter by taking $M$=$\hbar$=$c$=$k_B$=$\omega$=$1$. Although ML parameter is assumed to have a small value, such as $\beta=0.005$, we take into account two other values,  $\beta = 0$ and $\beta=0.5$, to discuss the OQM and asymptotic limits.  We show the thermodynamic functions in figs. (\ref{HFE}), (\ref{Ent}), (\ref{Int}), and (\ref{SpH}). In each graph, we illustrate the thermal properties with three different values of the cyclotron frequency. To be more specific, we take $\widetilde{\omega}=0$ value to investigate the case where the external magnetic field does not exist. We employ $\widetilde{\omega}=\omega$ and $\widetilde{\omega}=10\omega$ values to represent to the weak and strong external magnetic fields, respectively.

In all thermodynamic functions, we observe a critical value of the temperature, $T_c$, where the thermal behaviors alter. This critical value depends on the $\beta$ and $\widetilde{\omega}$ parameters and can be calculated from the roots of Eq. (\ref{MLPF}). Below this critical temperature, in the $0<T<T_c$ interval, the usual and well-know thermal behaviors are being violated. This can be seen as evidence of an existing fundamental low limit to measure the temperature. However, when $\beta=0$, it is still possible to define a lower limit to the temperature. We get this critical temperature out of  Eq. (\ref{OQMLZ}) as follows.
\begin{eqnarray}
  T_c &=& \sqrt{\frac{\hbar Mc^2}{k_B^2} \Big(\Omega-\omega\Big)}. \label{TC}
\end{eqnarray}
We note that $T_c$ is equal to zero if and only if $\widetilde{\omega}=0$. When we take $\widetilde{\omega}=\omega$ and $\widetilde{\omega}=10\omega$, we calculate the
critical temperature as $0.643$ and $3.008$, respectively. Therefore, we conclude that the existing of a lower temperature limit value is associated with the Euler-MacLaurin approximation which is employed in our solution. In the literature, it is noted that the Euler-MacLaurin approximation is appropriate for high-temperatures \cite{Wang_et_al_2015} and the low-temperature behaviors can be examined by adopting the method based on the Hurwitz zeta function \cite{Boumali_s_2015}.

\begin{figure}[!htb]
\includegraphics[width=0.80\linewidth]{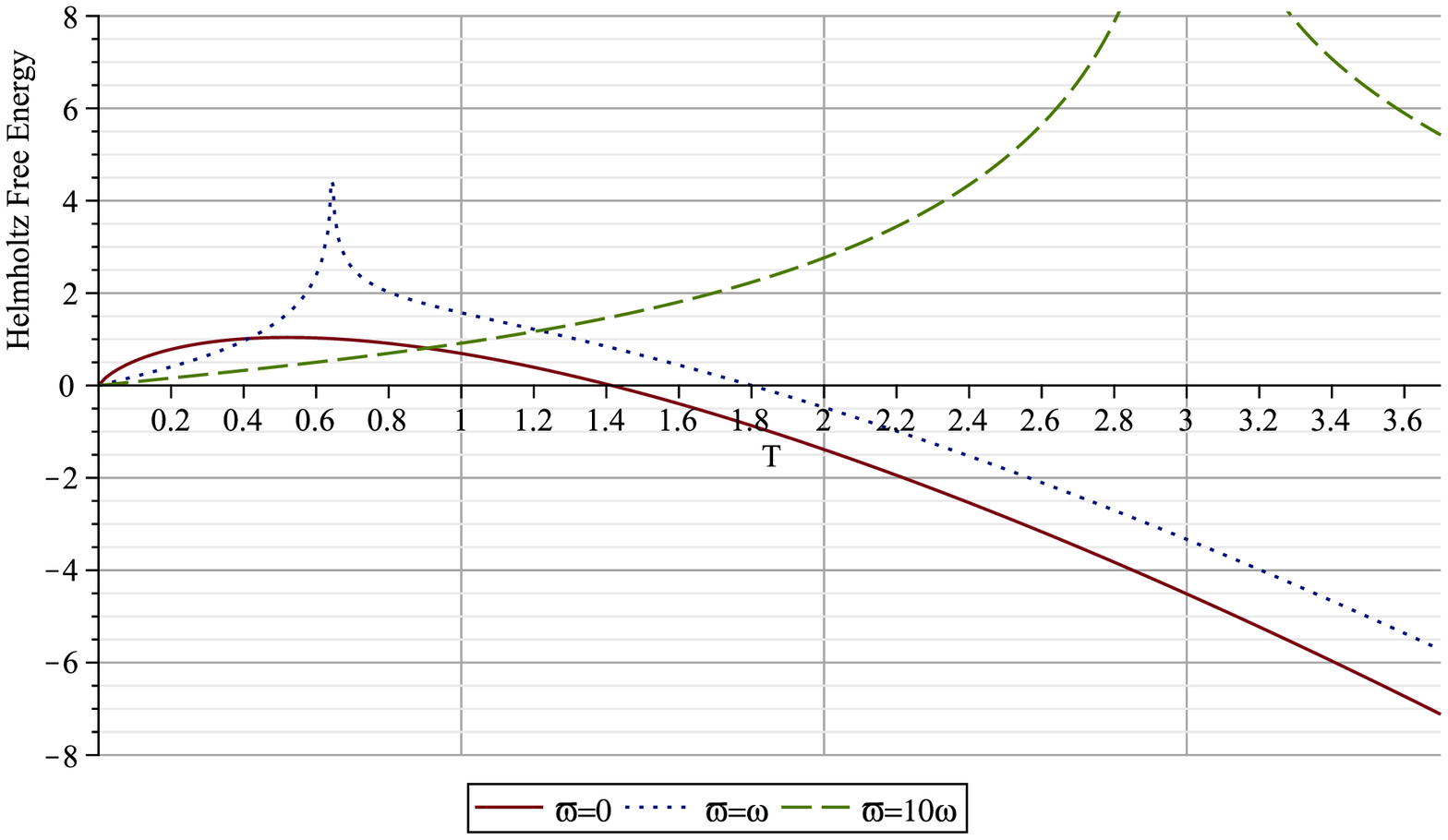}
\caption{Low temperature behavior of the Helmholtz free energy function in the OQM limit.} \label{LLHFE}
\end{figure}

Among the thermoynamic functions, first we examine the Helmholtz free energy. Before we proceed with the high-temperature behavior, we present the  low-temperature behavior in Fig. (\ref{LLHFE}) to illustrate the anomalies. The observed singularities around the $T_c\approx 0.6$ and $T_c\approx3.0$ values confirm our prediction. Then, in Fig. (\ref{HFE}) we demonstrate the Helmholtz free energy with a high-temperature range. We observe that all curves decrease monotonically via the temperature. Moreover, in larger ML parameters, the decreases in the Helmholtz free energy increase with equal temperature increases. For a fixed value of the temperature greater than the critical temperature, Helmholtz free energy has a higher value for the stronger external magnetic field.

Second, we investigate the entropy function. Although, we illustrate the entropy function together with low and high-temperature ranges in Fig. (\ref{Ent}), we interpret high-temperature behaviors. We observe a rapid change in entropy following the critical temperature value.  Then, the entropy functions start to grow more slowly.  For a fixed value of high temperature, in all cases, we observe the entropy function has a smaller value for higher external magnetic field values.

Next, we analyze the internal energy function of the system and present them in Fig. (\ref{Int}). In the OQM limit when the external magnetic field does not exist, we observe a linear mean energy function as is envisaged in Eq. (\ref{OQMNMFU}). Moreover, in the same limit, near the vicinity of the critical temperature, the internal energy function decreases rapidly, then it starts to increase like the other one. When we compare the total amount of energy increases in each case, we find that the $\beta=0$ is the lowest one. Except for the $\beta=0.005,\; \widetilde{\omega}=10\omega$  case, we observe very similar characteristic behaviors in this thermal quantity.

Finally, we demonstrate the specific heat functions in Fig. (\ref{SpH}). In the ordinary quantum mechanic limit, the specific heat function converges the half value of the other cases. We observe that the strength of the magnetic field plays a role in the peak value of the specific heat function. As the field strength increases, the peak value is obtained with a smaller value at a higher temperature.

\section{Conclusion} \label{Concl}
In this article, we studied the two dimensional DKP oscillator under an external homogeneous magnetic field in the GUP scenario. Initially, we decided to review the problem in the absence of the ML. We obtained an exact solution in the momentum representation in the spin-zero sector and verified the solutions with the ones in the literature. Then, we examined the spin-one sector. In some papers, we have seen that the authors made an ansatz that converts investigated solutions of the spin-one case equivalent to the spin-zero case's solution. Instead of that ansatz, we solved the spin-one case with an alternative approach in the non-relativistic limit. Then, we revisited the DKP oscillator problem in the ML formalism in momentum space. We derived an analytic expression for the energy eigenvalue function which depends on the ML parameter. We showed that in the absence of the ML parameter, the results become identical with the OQM ones in the non-relativistic limit. Finally, we studied the thermal properties by obtaining the Helmholtz free energy, entropy, mean energy and specific heat functions in the high-temperature limit. We obtained a lower limit critical temperature value for which the Euler-MacLaurin approach starts to be valid. We plotted the thermodynamic functions and discussed the effect of the external magnetic field on the thermal quantities of the DKP oscillator for values above this limit value.  We observed that the Helmholtz free energy function decreases monotonically with the increase of the temperature. We noticed that these decreases increase with equal temperature increases in the larger ML values. We found out an expeditious increase in the entropy functions just after the critical temperature value.  In the high-temperature values, the entropy functions grew slower than before.  For a fixed value of high temperature, in all cases with higher external magnetic field values, we observed the entropy function has a smaller value. In the analysis of the internal energy functions of the system, we derived a similar linear increase in all cases after the critical temperature value. We noticed that the strength of the external magnetic field affects the peak value of the specific heat function. When we considered a stronger external magnetic field, we observed a smaller peak value at a higher temperature.

\section*{Acknowledgment}
We like to thank for the support and encouragement of M. Heddar. We thank to E. Pehlivan, and  E. G\"{u}mr\"{u}k\c{c}\"{u}o\u{g}lu for the proofreading. BCL and JK are grateful to the support given by the Internal Project of Excellent Research of the Faculty of Science of University Hradec Kr\'{a}lov\'{e}, “2019/2216”. Finally, the authors thank the kind referee for his/her comments and valuable suggestions that leads a significant improvement of the quality of the manuscript.
%One of the authors, BCL, was partially supported by the Turkish Science and Research Council (TUBITAK).

%\end{document}

\newpage

\newpage

\newpage

\newpage

\newpage

\end{document}